\documentclass[12pt]{article}
     \usepackage{amssymb}
     \usepackage[reqno]{amsmath}
     \usepackage{pstricks}
\def\branecolor{black}
\def\backgroundcolor{black}
\def\flowcolor{black}

\def\parafermionbranemod{
\pspicture(-2.5,-2.5)(2.5,2.5)
\pscircle[linecolor=\backgroundcolor,linewidth=1pt](0,0){2}
\psset{linecolor=\branecolor,linewidth=2pt}
\qline(1.532088886,1.285575219)(-1,1.732050807)
\psset{dotsize=3pt 0}
\psdots(2,0)(1.532088886,1.285575219)(.3472963546,1.969615506)(-1,1.732050807)(-1.879385242,.6840402852)(-1.879385242,-.6840402866)(-1,-1.732050807)(.3472963600,-1.969615505)(1.532088888,-1.285575217)
\psset{linecolor=black,linewidth=1pt}
\qline(1.532088886,1.285575219)(0,0)
\qline(-1,1.732050807)(0,0)
\psarc(0,0){1}{40}{120}
\psline[linestyle=dashed](0,0)(.2660444426,1.508813014)
\psline[linestyle=dashed](0,0)(2,0)
\psarc[linestyle=dashed](0,0){0.7}{0}{80}
\rput(0.7,-0.6){$\pi \frac{L'}{k}$}
\psline[linewidth=0.5pt](0.5,0.15)(0.6,-0.3)
\rput(-1,0.3){$2 \pi \frac{L}{k}$}
\psline[linewidth=0.5pt](-0.6,0.3)(-0.1,0.7)
\endpspicture
}
\def\paraprocI{
\pspicture(-6,-2.5)(6,2.5)
\rput(-3.5,0){
\pscircle[linecolor=\backgroundcolor,linewidth=1pt](0,0){2}
\psset{linecolor=\branecolor,linewidth=2pt}
\qline(2,0)(.3472963546,1.969615506)
\qline(1.532088886,1.285575219)(-1,1.732050807)}
\rput(3.5,0){
\pscircle[linecolor=\backgroundcolor,linewidth=1pt](0,0){2}
\psset{linecolor=\branecolor,linewidth=2pt}
\qline(2,0)(-1,1.732050807)
\qline(1.532088886,1.285575219)(.3472963546,1.969615506)
}
\psline[linewidth=1.5pt,linecolor=\flowcolor]{->}(-0.8,0)(0.8,0)
\endpspicture
}
\def\parafermiontwbrane{
\pspicture(-2.5,-2.5)(2.5,2.5)
\pscircle[linecolor=\backgroundcolor,linewidth=1pt](0,0){2}
\psset{linecolor=\branecolor,linewidth=2pt}
\pscircle[linewidth=1pt,fillstyle=hlines,hatchsep=1pt](0,0){.5}
\psset{dotsize=3pt 0}
\psdots(2,0)(1.532088886,1.285575219)(.3472963546,1.969615506)(-1,1.732050807)(-1.879385242,.6840402852)(-1.879385242,-.6840402866)(-1,-1.732050807)(.3472963600,-1.969615505)(1.532088888,-1.285575217)
\psset{linecolor=black,linewidth=1pt}
\endpspicture
}
\def\paratwproc{
\pspicture(-6,-2.5)(6,2.5)
\rput(-3.5,0){
\pscircle[linecolor=\backgroundcolor,linewidth=1pt](0,0){2}
\psset{linecolor=\branecolor,linewidth=2pt}
\psset{dotsize=5pt 0}
\psdots(0,0.1)(0,-0.1)}
\rput(3.5,0){
\pscircle[linecolor=\backgroundcolor,linewidth=1pt](0,0){2}
\psset{linecolor=\branecolor,linewidth=2pt}
\pscircle[linewidth=1pt,fillstyle=hlines,hatchsep=1pt](0,0){.5}}
\psline[linewidth=1.5pt,linecolor=\flowcolor]{->}(-0.8,0)(0.8,0)
\endpspicture
}

\def\parafermionprocess{
\pspicture(-6,-2.5)(6,2.5)
\rput(-3.5,0){
\pscircle[linecolor=\backgroundcolor,linewidth=1pt](0,0){2}
\psset{linecolor=\branecolor,linewidth=2pt}
\psset{dotsize=5pt 0}
\psdots(2,0)(1.532088886,1.285575219)(.3472963546,1.969615506)(-1,1.732050807)}
\rput(3.5,0){
\pscircle[linecolor=\backgroundcolor,linewidth=1pt](0,0){2}
\psset{linecolor=\branecolor,linewidth=2pt}
\qline(2,0)(-1,1.732050807)}
\psline[linewidth=1.5pt,linecolor=\flowcolor]{->}(-0.8,0)(0.8,0)
\endpspicture
}

\def\minimalmodelgeometryfinitek{
\pspicture(-6.7,-2)(6.7,2)
\psset{linecolor=\backgroundcolor}
\qline(-6.5,-1.5)(-1,-1.5)
\qline(0.5,-1.5)(5.5,-1.5)
\qline(-6.5,1.5)(-1,1.5)
\qline(0.5,1.5)(5.5,1.5)
\qline(-6.5,1.5)(-6.5,-1.5)
\psset{linestyle=dashed}
\qline(5.5,-1.5)(7,-1.5)
\qline(-1,-1.5)(0.5,-1.5)
\qline(5.5,1.5)(7,1.5)
\qline(-1,1.5)(0.5,1.5)
\psset{linestyle=solid}
\def\myellipse{\scalebox{0.2 1}{\psarc[linestyle=dashed](0,0){1.5}{270}{90}\psarc(0,0){1.5}{90}{270}}}
\rput(-4.2625,0){\myellipse}
\rput(-2.025,0){\myellipse}
\rput(1,0){\myellipse}
\rput(5,0){\myellipse}
\psline{->}(-7.3,-2.1)(-7.3,1.3)
\psline{->}(-7.3,-2.1)(7.3,-2.1)
\psline{->}(-7.3,-2.1)(-6.7,-1.6)
\rput(-7.5,1.2){$z$}
\rput(7,-1.9){$x$}
\rput(-6.95,-1.5){$y$}
\psset{linecolor=\branecolor,hatchcolor=\branecolor}
\pscircle*(-6.5,1.5){0.1}
\pscircle*(-4,-1.5){0.1}
\pscircle*(-1.5,1.5){0.1}
\rput(-2.025,0){\scalebox{0.2
1}{\psline[linewidth=2pt](-1.469693846,0.3)(1.299038106,0.75)}}
\rput(-4.2625,0){\scalebox{0.2
1}{\psline[linewidth=2pt](-.9508659879,-1.160109423)(1.269321525,-.7992639520)}}
\rput(1,0){{\pspolygon[hatchsep=1pt,fillstyle=hlines](.2939387692,-0.3)(-.2598076212,-0.75)(3.740192378,-0.75)(4.293938770,-0.3)}}
\rput(-6.5,1.8){$(0,0,0)$}
\rput(-4,-1.8){$(1,1,0)$}
\rput(-3.3,-0.8){{$(0,1,1)$}}
\rput(-1.5,1.8){$(2,0,0)$}
\rput(-1.1,0.6){$(0,0,2)$}
\rput(3,0){$(L_{1},L_{2},L')$}
\psline{<->}(5.5,-1.5)(5.5,-0.525)
\rput(6.1,-1){$\displaystyle\frac{L'}{k+1}$}
\endpspicture
}

\def\minimalmodelDA{
\pspicture(-5,-2)(5,2)
\def\myellipse{\scalebox{0.2 1}{\psarc[linestyle=dashed](0,0){.8}{270}{90}\psarc[linestyle=dashed](0,0){.8}{90}{270}}}
\psset{linecolor=\backgroundcolor}
\psline(-5,.8)(-5,-.8)(5,-.8)(5,.8)(-5,.8)
\psset{linewidth=2pt,linecolor=\flowcolor}
\psline{->}(0,0)(4.5,0)
\psline{->}(0,0)(-4.5,0)
\myellipse
\endpspicture
}

\def\minmodflowsDAI{
\pspicture(-6.7,-2.5)(6.7,2.5)
 
\def\myellipse{\scalebox{0.2 1}{\psarc[linestyle=dashed](0,0){.8}{270}{90}\psarc[linestyle=dashed](0,0){.8}{90}{270}}}
\def\mypolygon{\pspolygon[hatchsep=1pt,fillstyle=hlines,hatchcolor=\branecolor](1,-0.5)(3.5,-0.5)(3.9,-0.125)(1.4,-0.125)}
\rput(-6.5,0){\psset{linecolor=\backgroundcolor}\qline(0,.8)(5,.8)\qline(0,-.8)(5,-.8)\qline(0,.8)(0,-.8)\rput(5,0){\myellipse}
\psset{linecolor=\branecolor}\mypolygon}

\psline[linewidth=1.5pt,linecolor=\flowcolor]{->}(-0.8,0.4)(0.8,0.7)
\psline[linewidth=1.5pt,linecolor=\flowcolor]{->}(-0.8,-0.4)(0.8,-0.7)
\rput(1.5,1.2){\psset{linecolor=\backgroundcolor}\qline(0,.8)(5,.8)\qline(0,-.8)(5,-.8)\qline(0,-.8)(0,.8)\rput(5,0){\myellipse}\psset{linecolor=\branecolor}\pscircle*(0.9,-.8){0.07}\pscircle*(2.45,-.8){0.07}\pscircle*(4,-.8){0.07}}
\rput(1.5,-1.2){\psset{linecolor=\backgroundcolor}\qline(0,.8)(5,.8)\qline(0,-.8)(5,-.8)\qline(0,-.8)(0,.8)\rput(5,0){\myellipse}\psset{linecolor=\branecolor}\pscircle*(1.675,.8){0.07}\pscircle*(3.225,.8){0.07}}
\endpspicture
}

\def\minmodflowsDAII{
\pspicture(-6.7,-2.5)(6.7,2.5)
 
\def\myellipse{\scalebox{0.2 1}{\psarc[linestyle=dashed](0,0){.8}{270}{90}\psarc[linestyle=dashed](0,0){.8}{90}{270}}}
\def\mypolygon{\pspolygon[hatchsep=1pt,fillstyle=hlines,hatchcolor=\branecolor](2.3,0.125)(4.82,0.125)(5.125,0.411175)(2.605,0.411175)}
\def\mypolygona{\pspolygon[hatchsep=1pt,fillstyle=vlines,hatchcolor=\branecolor](3.65,0.125)(4.82,0.125)(5.125,0.411175)(3.955,0.411175)}
\rput(-6.5,0){\psset{linecolor=\backgroundcolor}\qline(0,.8)(5,.8)\qline(0,-.8)(5,-.8)\qline(0,.8)(0,-.8)\rput(5,0){\myellipse}
\psset{linecolor=\branecolor}\mypolygon\mypolygona}

\psline[linewidth=1.5pt,linecolor=\flowcolor]{->}(-0.8,0.4)(0.8,0.7)
\psline[linewidth=1.5pt,linecolor=\flowcolor]{->}(-0.8,-0.4)(0.8,-0.7)
\rput(1.5,1.2){\psset{linecolor=\backgroundcolor}\qline(0,.8)(5,.8)\qline(0,-.8)(5,-.8)\qline(0,-.8)(0,.8)\rput(5,0){\myellipse}\psset{linecolor=\branecolor}\pscircle*(1.9,.8){0.07}\pscircle*(3.45,.87){0.07}\pscircle*(3.45,.73){0.07}\pscircle*(4.93,.73){0.07}\pscircle*(4.93,.87){0.07}\pscircle*(5.07,.73){0.07}\pscircle*(5.07,.87){0.07}\rput(4.97,1.05){+ --}
}
\rput(1.5,-1.2){\psset{linecolor=\backgroundcolor}\qline(0,.8)(5,.8)\qline(0,-.8)(5,-.8)\qline(0,-.8)(0,.8)\rput(5,0){\myellipse}\psset{linecolor=\branecolor}\pscircle*(2.675,-.8){0.07}\pscircle*(4.225,-.73){0.07}\pscircle*(4.225,-.87){0.07}}
\endpspicture
}
\def\minmodflowsDAIII{
\pspicture(-6.7,-2.5)(6.7,2.5)
 
\def\myellipse{\scalebox{0.2 1}{\psarc[linestyle=dashed](0,0){.8}{270}{90}\psarc[linestyle=dashed](0,0){.8}{90}{270}}}
\def\mypolygon{\pspolygon[hatchsep=1pt,fillstyle=hlines,hatchcolor=\branecolor](3.65,0.125)(4.82,0.125)(5.125,0.411175)(3.955,0.411175)}
\def\mypolygona{\pspolygon[hatchsep=1pt,fillstyle=vlines,hatchcolor=\branecolor](3.65,0.125)(4.82,0.125)(5.125,0.411175)(3.955,0.411175)}
\rput(-6.5,0){\psset{linecolor=\backgroundcolor}\qline(0,.8)(5,.8)\qline(0,-.8)(5,-.8)\qline(0,.8)(0,-.8)\rput(5,0){\myellipse}
\psset{linecolor=\branecolor}\mypolygon\mypolygona\rput(3.5,.3){+}}

\psline[linewidth=1.5pt,linecolor=\flowcolor]{->}(-0.8,0.4)(0.8,0.7)
\psline[linewidth=1.5pt,linecolor=\flowcolor]{->}(-0.8,-0.4)(0.8,-0.7)
\rput(1.5,1.2){\psset{linecolor=\backgroundcolor}\qline(0,.8)(5,.8)\qline(0,-.8)(5,-.8)\qline(0,-.8)(0,.8)\rput(5,0){\myellipse}\psset{linecolor=\branecolor}\pscircle*(3.45,.8){0.07}\pscircle*(5,.8){0.07}\rput(5,1.05){--}
}
\rput(1.5,-1.2){\psset{linecolor=\backgroundcolor}\qline(0,.8)(5,.8)\qline(0,-.8)(5,-.8)\qline(0,-.8)(0,.8)\rput(5,0){\myellipse}\psset{linecolor=\branecolor}\pscircle*(4.225,-.8){0.07}}
\endpspicture
}
\def\minimalmodelAD{
\pspicture(-5,-2)(5,2)
\def\myellipse{\scalebox{0.2 1}{\psarc[linestyle=dashed](0,0){.8}{270}{90}\psarc[linestyle=dashed](0,0){.8}{90}{270}}}
\psset{linecolor=\backgroundcolor}
\psline(-5,.8)(-5,-.8)(5,-.8)(5,.8)(-5,.8)
\pscircle*(0,0){0.1}
\psset{linewidth=2pt,linecolor=\flowcolor}
\psline{->}(0,0)(4.5,.72)
\psline{->}(0,0)(-4.5,-.72)
\psline{->}(0,0)(4.5,-.72)
\psline{->}(0,0)(-4.5,.72)
\myellipse
\endpspicture
}
\def\minmodflowsADI{
\pspicture(-6.7,-2.5)(6.7,2.5)
 
\def\myellipse{\scalebox{0.2 1}{\psarc[linestyle=dashed](0,0){.8}{270}{90}\psarc[linestyle=dashed](0,0){.8}{90}{270}}}
\def\mypolygon{\pspolygon[hatchsep=1pt,fillstyle=hlines,hatchcolor=\branecolor](1.9125,0.125)(4.82,0.125)(5.125,0.411175)(2.2175,0.411175)}
\def\mypolygona{\pspolygon[hatchsep=1pt,fillstyle=vlines,hatchcolor=\branecolor](4.075,-.411175)(4.82,-0.411175)(5.125,-0.125)(4.3775,-0.125)}
\rput(-6.5,0){\psset{linecolor=\backgroundcolor}\qline(0,.8)(5,.8)\qline(0,-.8)(5,-.8)\qline(0,.8)(0,-.8)\rput(5,0){\myellipse}
\psset{linecolor=\branecolor}\mypolygon\mypolygona}

\psline[linewidth=1.5pt,linecolor=\flowcolor]{->}(-0.8,0.4)(0.8,0.7)
\psline[linewidth=1.5pt,linecolor=\flowcolor]{->}(-0.8,-0.4)(0.8,-0.7)
\rput(1.5,1.2){\psset{linecolor=\backgroundcolor}\qline(0,.8)(5,.8)\qline(0,-.8)(5,-.8)\qline(0,-.8)(0,.8)\rput(5,0){\myellipse}\psset{linecolor=\branecolor}\pscircle*(1.5125,.8){0.07}\pscircle*(3.0625,.8){0.07}\pscircle*(4.6125,.8){0.07}\pscircle*(3.8375,-.8){0.07}
}
\rput(1.5,-1.2){\psset{linecolor=\backgroundcolor}\qline(0,.8)(5,.8)\qline(0,-.8)(5,-.8)\qline(0,-.8)(0,.8)\rput(5,0){\myellipse}\psset{linecolor=\branecolor}\pscircle*(2.2875,-.8){0.07}\pscircle*(3.8375,-.8){0.07}\pscircle*(4.6125,.8){0.07}}
\endpspicture
}
\def\minmodflowsADII{
\pspicture(-6.7,-1.2)(6.7,1.2)
 
\def\myellipse{\scalebox{0.2 1}{\psarc[linestyle=dashed](0,0){.8}{270}{90}\psarc[linestyle=dashed](0,0){.8}{90}{270}}}
\def\mypolygon{\pspolygon[hatchsep=1pt,fillstyle=hlines,hatchcolor=\branecolor](1.7524509,-.1475491)(4.84274,-.1475491)(5.157255,0.1475491)(2.0475491,0.1475491)}
\def\mypolygona{\pspolygon[hatchsep=1pt,fillstyle=vlines,hatchcolor=\branecolor](1.7524509,-.1475491)(4.84274,-.1475491)(5.157255,0.1475491)(2.0475491,0.1475491)}
\rput(-6.5,0){\psset{linecolor=\backgroundcolor}\qline(0,.8)(5,.8)\qline(0,-.8)(5,-.8)\qline(0,.8)(0,-.8)\rput(5,0){\myellipse}
\psset{linecolor=\branecolor}\mypolygon\mypolygona\rput(1.6,-.1){$\begin{array}{c}+\\[-1mm]
 - \end{array}$}}

\psline[linewidth=1.5pt,linecolor=\flowcolor]{->}(-0.8,0)(0.8,0)
\rput(1.5,0){\psset{linecolor=\backgroundcolor}\qline(0,.8)(5,.8)\qline(0,-.8)(5,-.8)\qline(0,-.8)(0,.8)\rput(5,0){\myellipse}\psset{linecolor=\branecolor}\pscircle*(1.5125,.8){0.07}\pscircle*(3.0625,.8){0.07}\pscircle*(4.6125,.8){0.07}\pscircle*(3.8375,-.8){0.07}\pscircle*(2.2875,-.8){0.07}
}
\endpspicture
}

\def\isingmodelflows{
\pspicture(-1,-1.5)(3,1.5)
\psset{linecolor=\backgroundcolor}\pspolygon(2,1)(0,1)(0,-1)(2,-1)\psset{linecolor=\branecolor}\pscircle*(0,1){0.1}\pscircle*(2,-1){0.1}\rput(1,0.35){\scalebox{0.2
1}{\psline[linewidth=2pt](.9797958973,-0.2)(-.8660254040,-0.5)}}\psset{linecolor=\flowcolor}\psline{->}(1.1,-0.1)(1.9,-0.85)\psline{->}(0.9,0.1)(0.1,0.85)

\endpspicture
}

\def\minmodsmallkprocesses{
\pspicture(-6.7,-3.6)(6.7,3.6)
\def\myellipse{\scalebox{0.2 1}{\psarc[linestyle=dashed](0,0){1.3}{270}{90}\psarc(0,0){1.3}{90}{270}}}
\def\mypolygon{\pspolygon[hatchsep=1pt,fillstyle=hlines,hatchcolor=\branecolor](0.7,-0.8)(3.8,-0.8)(4.3,-0.2)(1.2,-0.2)}

\rput(-6.5,0){\psset{linecolor=\backgroundcolor}\qline(0,1.3)(5,1.3)\qline(0,-1.3)(5,-1.3)\myellipse\rput(5,0){\myellipse}
\psset{linecolor=\branecolor}\mypolygon}

\psline[linewidth=1.5pt,linecolor=\flowcolor]{->}(-0.8,0.5)(0.8,1.5)
\psline[linewidth=1.5pt,linecolor=\flowcolor]{->}(-0.8,-0.5)(0.8,-1.5)

\rput(1.5,2){\psset{linecolor=\backgroundcolor}\qline(0,1.3)(5,1.3)\qline(0,-1.3)(5,-1.3)\myellipse\rput(5,0){\myellipse}\psset{linecolor=\branecolor}\pscircle*(1.175,-1.3){0.1}\pscircle*(2.725,-1.3){0.1}\pscircle*(4.275,-1.3){0.1}}

\rput(1.5,-2){\psset{linecolor=\backgroundcolor}\qline(0,1.3)(5,1.3)\qline(0,-1.3)(5,-1.3)\myellipse\rput(5,0){\myellipse}\psset{linecolor=\branecolor}\pscircle*(0.4,1.3){0.1}\pscircle*(1.95,1.3){0.1}\pscircle*(3.5,1.3){0.1}}
\endpspicture
}
\def\minmodsmallkprocessesII{
\pspicture(-6.7,-3.6)(6.7,3.6)
\def\myellipse{\scalebox{0.2 1}{\psarc[linestyle=dashed](0,0){1.3}{270}{90}\psarc(0,0){1.3}{90}{270}}}
\def\mypolygon{\pspolygon[hatchsep=1pt,fillstyle=hlines,hatchcolor=\branecolor](0.925,-0.8)(3.8,-0.8)(4.3,-0.2)(1.425,-0.2)}

\rput(-6.5,0){\psset{linecolor=\backgroundcolor}\qline(0,1.3)(5,1.3)\qline(0,-1.3)(5,-1.3)\myellipse\rput(5,0){\myellipse}
\psset{linecolor=\branecolor}\mypolygon}

\psline[linewidth=1.5pt,linecolor=\flowcolor]{->}(-0.8,0.5)(0.8,1.5)
\psline[linewidth=1.5pt,linecolor=\flowcolor]{->}(-0.8,-0.5)(0.8,-1.5)

\rput(1.5,2){\psset{linecolor=\backgroundcolor}\qline(0,1.3)(5,1.3)\qline(0,-1.3)(5,-1.3)\myellipse\rput(5,0){\myellipse}\psset{linecolor=\branecolor}\pscircle*(1.175,-1.3){0.1}\pscircle*(2.725,-1.3){0.1}\pscircle*(4.275,-1.3){0.1}}

\rput(1.5,-2){\psset{linecolor=\backgroundcolor}\qline(0,1.3)(5,1.3)\qline(0,-1.3)(5,-1.3)\myellipse\rput(5,0){\myellipse}\psset{linecolor=\branecolor}\pscircle*(1.95,1.3){0.1}\pscircle*(3.5,1.3){0.1}}
\endpspicture
}
\def\minmodsmallkprocessesIII{
\pspicture(-6.7,-3.6)(6.7,3.6)
 
\def\myellipse{\scalebox{0.2 1}{\psarc[linestyle=dashed](0,0){1.3}{270}{90}\psarc(0,0){1.3}{90}{270}}}
\def\mypolygon{\pspolygon[hatchsep=1pt,fillstyle=hlines,hatchcolor=\branecolor](0.7,-0.8)(3.8,-0.8)(4.3,-0.2)(1.2,-0.2)}
\def\mypolygonII{\pspolygon[hatchsep=1pt,fillstyle=hlines,hatchcolor=\branecolor](0.9,-0.8)(3.6,-0.8)(4.1,-0.2)(1.4,-0.2)}

\rput(-6.5,0){\psset{linecolor=\backgroundcolor,linewidth=1pt}\qline(0,1.3)(5,1.3)\qline(0,-1.3)(5,-1.3)\myellipse\rput(5,0){\myellipse}
\psset{linecolor=\branecolor,linewidth=2pt}\qline(0.7,-0.5)(1.2,0.1)\qline(2.25,-0.5)(2.75,0.1)\qline(3.8,-0.5)(4.3,0.1)}

\psline[linewidth=1.5pt,linecolor=\flowcolor]{->}(-0.8,0.5)(0.8,1.5)
\psline[linewidth=1.5pt,linecolor=\flowcolor]{->}(-0.8,-0.5)(0.8,-1.5)

\rput(1.5,2){\psset{linecolor=\backgroundcolor,linewidth=1pt}\qline(0,1.3)(5,1.3)\qline(0,-1.3)(5,-1.3)\myellipse\rput(5,0){\myellipse}\psset{linecolor=\branecolor}\mypolygon}
\rput(1.5,-2){\psset{linecolor=\backgroundcolor,linewidth=1pt}\qline(0,1.3)(5,1.3)\qline(0,-1.3)(5,-1.3)\myellipse\rput(5,0){\myellipse}\psset{linecolor=\branecolor}\rput(0,1){\mypolygonII}}

\endpspicture
}

\def\tricriticalisingmodel{
\pspicture(-6,-3)(6,3)
\pspolygon[linecolor=\backgroundcolor](-5,-2.5)(-5,2.5)(5,2.5)(5,-2.5)
\psset{linecolor=\branecolor,hatchcolor=\branecolor}
\pscircle*(5,2.5){0.1} 
\pscircle*(-5,2.5){0.1}
\pscircle*(0,-2.5){0.1}
\pspolygon[hatchsep=2pt,fillstyle=hlines](-4,0.5)(3.3,0.5)(4,1.5)(-3.3,1.5)
\psline[linewidth=2pt](-2.2,-1.5)(-1.5,-0.5) 
\psline[linewidth=2pt](1.5,-1.5)(2.2,-0.5)   
\psset{linecolor=\flowcolor}
\psline{->}(-1.8,-1.1)(-0.2,-2.3) 
\psline{->}(1.6,-1.2)(0.2,-2.3)   
\psline{->}(0,0.3)(0,-2.1)        
\psline(0,1.7)(0,1.9)                                       
\psline[linearc=0.3]{->}(0,1.9)(0,2.2)(-4.5,2.2)(-4.7,2.35) 
\psline[linearc=0.3]{->}(0,1.9)(0,2.2)(4.5,2.2)(4.7,2.35)   
\psline[linearc=0.5]{->}(-2.0,-0.95)(-4.3,0.2)(-4.8,2.25)
\psline[linearc=0.5]{->}(2.0,-0.95)(4.3,0.2)(4.8,2.25) 
\rput(0.6,1.85){$\scriptstyle{+\,}\textstyle\phi_{13}$}
\rput(-1.5,-1.85){$\scriptstyle{+\,}\textstyle\phi_{13}$}
\rput(3.5,-0.6){$\scriptstyle{+\,}\textstyle\phi_{13}$}
\rput(0.6,0){$\scriptstyle{-\,}\textstyle\phi_{13}$}
\rput(1.25,-2){$\scriptstyle{-\,}\textstyle\phi_{13}$}
\rput(-3.5,-0.6){$\scriptstyle{-\,}\textstyle\phi_{13}$}
\psset{linewidth=2pt,linestyle=dotted}
\psline{->}(-3.8,1.0)(-4.6,2.1)   
\psline{->}(4,1.5)(4.6,2.1)     
\rput(-3.85,1.8){$\scriptstyle{-\,}\textstyle\phi_{33}$}
\rput(3.7,1.85){$\scriptstyle{+\,}\textstyle\phi_{33}$}
\rput(-5,2.8){$\scriptstyle (1,1)$}
\rput(5,2.8){$\scriptstyle (3,1)$}
\rput(0,-2.8){$\scriptstyle (2,1)$}
\rput(-1.2,-0.3){$\scriptstyle (1,2)$}
\rput(2.3,-0.3){$\scriptstyle (1,3)$}
\rput(-2,1){\psframebox*{$\scriptstyle (2,2)$}}
\endpspicture
}

\def\pottsgeometry{
\pspicture(-6,-2)(6,2)
\rput(-4.5,0){
\pscircle[linecolor=\backgroundcolor,linewidth=1pt](0,0){1.5}
\psset{linecolor=\branecolor,linewidth=2pt}
\psset{dotsize=5pt 0}
\psdots(1.5,0)(-0.75,1.299038106)(-0.75,-1.299038106)(0,0)
\rput(1.75,0){$A$}
\rput(-1,1.5){$B$}
\rput(-1,-1.5){$C$}
\rput(-0.6,0){`free'}
}
\rput(0,0){
\pscircle[linecolor=\backgroundcolor,linewidth=1pt](0,0){1.5}
\psset{linecolor=\branecolor,linewidth=1.5pt}
\pspolygon(1.5,0)(-0.75,1.299038106)(-0.75,-1.299038106)
\rput(.65,.9){$AB$}
\rput(-1.1,0){$BC$}
\rput(.6,-.85){$AC$}
}
\rput(4.5,0){
\pscircle[linecolor=\backgroundcolor,linewidth=1pt](0,0){1.5}
\psset{linecolor=\branecolor,linewidth=1pt,hatchcolor=\branecolor}
\pscircle[fillstyle=hlines,hatchsep=4pt](0,0){1.05}
\rput(0,-0.5){\psframebox*{`new'}}
}
\endpspicture
}

\def\pottsAundB{\parbox{1cm}{\pspicture(-0.5,-0.5)(0.5,0.5)
\scalebox{0.5}{
\pscircle[linecolor=\backgroundcolor,linewidth=0.5pt](0,0){1}
\psset{dotsize=5pt 0}
\psset{linecolor=\branecolor,linewidth=1.5pt} 
\psdots(1,0)(-0.5,.866025404)
}
\endpspicture}}
\def\pottsAB{\parbox{1cm}{\pspicture(-0.5,-0.5)(0.5,0.5)
\scalebox{0.5}{
\pscircle[linecolor=\backgroundcolor,linewidth=0.5pt](0,0){1}
\psset{dotsize=5pt 0}
\psset{linecolor=\branecolor,linewidth=1.5pt} 
\psline(1,0)(-0.5,.866025404)
}
\endpspicture}}
\def\pottsABundBC{\parbox{1cm}{\pspicture(-0.5,-0.5)(0.5,0.5)
\scalebox{0.5}{
\pscircle[linecolor=\backgroundcolor,linewidth=0.5pt](0,0){1}
\psset{dotsize=5pt 0}
\psset{linecolor=\branecolor,linewidth=1.5pt} 
\psline(1,0)(-0.5,.866025404)(-0.5,-.866025404)
}
\endpspicture}}
\def\pottsBundAC{\parbox{1cm}{\pspicture(-0.5,-0.5)(0.5,0.5)
\scalebox{0.5}{
\pscircle[linecolor=\backgroundcolor,linewidth=0.5pt](0,0){1}
\psset{dotsize=5pt 0}
\psset{linecolor=\branecolor,linewidth=1.5pt} 
\psline(1,0)(-0.5,-.866025404)
\psdots(-0.5,.866025404)
}
\endpspicture}}
\def\pottsfreefree{\parbox{1cm}{\pspicture(-0.5,-0.5)(0.5,0.5)
\scalebox{0.5}{
\pscircle[linecolor=\backgroundcolor,linewidth=0.5pt](0,0){1}
\psset{dotsize=5pt 0}
\psset{linecolor=\branecolor,linewidth=1.5pt} 
\psdots(-0.1,0)(0.1,0)
}
\endpspicture}}
\def\pottsfree{\parbox{1cm}{\pspicture(-0.5,-0.5)(0.5,0.5)
\scalebox{0.5}{
\pscircle[linecolor=\backgroundcolor,linewidth=0.5pt](0,0){1}
\psset{dotsize=5pt 0}
\psset{linecolor=\branecolor,linewidth=1.5pt} 
\psdots(0,0)
}
\endpspicture}}
\def\pottsnew{\parbox{1cm}{\pspicture(-0.5,-0.5)(0.5,0.5)
\scalebox{0.5}{
\pscircle[linecolor=\backgroundcolor,linewidth=0.5pt](0,0){1}
\psset{dotsize=5pt 0}
\psset{linecolor=\branecolor,linewidth=1.5pt} 
\pscircle[fillstyle=hlines,hatchsep=4pt,hatchcolor=\branecolor,linewidth=0.5pt](0,0){0.7}
}
\endpspicture}}
\def\pottsnewnew{\parbox{1cm}{\pspicture(-0.5,-0.5)(0.5,0.5)
\scalebox{0.5}{
\pscircle[linecolor=\backgroundcolor,linewidth=0.5pt](0,0){1}
\psset{dotsize=5pt 0}
\psset{linecolor=\branecolor,linewidth=1.5pt} 
\rput(0.06,0){\pscircle[fillstyle=hlines,hatchsep=4pt,hatchcolor=\branecolor,linewidth=0.5pt](0,0){0.7}}
\rput(-0.06,0){\pscircle[fillstyle=hlines,hatchsep=4pt,hatchcolor=\branecolor,linewidth=0.5pt](0,0){0.7}}
}
\endpspicture}}
\def\pottsA{\parbox{1cm}{\pspicture(-0.5,-0.5)(0.5,0.5)
\scalebox{0.5}{
\pscircle[linecolor=\backgroundcolor,linewidth=0.5pt](0,0){1}
\psset{dotsize=5pt 0}
\psset{linecolor=\branecolor,linewidth=1.5pt} 
\psdots(1,0)
}
\endpspicture}}
\def\pottsB{\parbox{1cm}{\pspicture(-0.5,-0.5)(0.5,0.5)
\scalebox{0.5}{
\pscircle[linecolor=\backgroundcolor,linewidth=0.5pt](0,0){1}
\psset{dotsize=5pt 0}
\psset{linecolor=\branecolor,linewidth=1.5pt} 
\psdots(-0.5,.866025404)
}
\endpspicture}}
\def\pottsfreenew{\parbox{1cm}{\pspicture(-0.5,-0.5)(0.5,0.5)
\scalebox{0.5}{
\pscircle[linecolor=\backgroundcolor,linewidth=0.5pt](0,0){1}
\psset{dotsize=5pt 0}
\psset{linecolor=\branecolor,linewidth=1.5pt} 
\psdots(0,0)
\pscircle[fillstyle=hlines,hatchsep=4pt,hatchcolor=\branecolor,linewidth=0.5pt](0,0){0.7}
}
\endpspicture}}
\def\pottsABC{\parbox{1cm}{\pspicture(-0.5,-0.5)(0.5,0.5)
\scalebox{0.5}{
\pscircle[linecolor=\backgroundcolor,linewidth=0.5pt](0,0){1}
\psset{dotsize=5pt 0}
\psset{linecolor=\branecolor,linewidth=1.5pt} 
\psdots(1,0)(-0.5,.866025404)(-0.5,-.866025404)
}
\endpspicture}}
\def\pottsAundBC{\parbox{1cm}{\pspicture(-0.5,-0.5)(0.5,0.5)
\scalebox{0.5}{
\pscircle[linecolor=\backgroundcolor,linewidth=0.5pt](0,0){1}
\psset{dotsize=5pt 0}
\psset{linecolor=\branecolor,linewidth=1.5pt} 
\psdots(1,0)
\psline(-0.5,.866025404)(-0.5,-.866025404)
}
\endpspicture}}
\def\pottsABundAC{\parbox{1cm}{\pspicture(-0.5,-0.5)(0.5,0.5)
\scalebox{0.5}{
\pscircle[linecolor=\backgroundcolor,linewidth=0.5pt](0,0){1}
\psset{dotsize=5pt 0}
\psset{linecolor=\branecolor,linewidth=1.5pt} 
\psline(-0.5,.866025404)(1,0)(-0.5,-.866025404)
}
\endpspicture}}
\def\pottsABundBCundCA{\parbox{1cm}{\pspicture(-0.5,-0.5)(0.5,0.5)
\scalebox{0.5}{
\pscircle[linecolor=\backgroundcolor,linewidth=0.5pt](0,0){1}
\psset{dotsize=5pt 0}
\psset{linecolor=\branecolor,linewidth=1.5pt} 
\psline(-0.5,.866025404)(1,0)(-0.5,-.866025404)(-0.5,.866025404)
}
\endpspicture}}

\def\pottsprocesses{
\pspicture(-6.5,-8,5)(6.5,4)
\scalebox{1.1}{
\rput(1,-7.5){\pottsA }
\rput(1,-5.5){\pottsAB }
\rput(-1,-4.5){\pottsfree }
\rput(1,-3){\pottsAundB }
\rput(1,-1){\pottsAundBC }
\rput(-1,0){\pottsnew }
\rput(-3,1){\pottsABC }
\rput(1,2){\pottsABundAC}
\rput(-1,3){\pottsABundBCundCA}

\psset{linewidth=1pt,linecolor=\flowcolor}
\psline{->}(1,-6.1)(1,-6.9)
\psline{->}(1,-3.6)(1,-4.9)
\psline{->}(-1,-0.6)(-1,-3.9)

\psline{->}(1,-1.6)(1,-2.4)
\psline{->}(-2.463343686,.7316718428)(-1.536656314,.2683281572)
\psline{->}(-2.794954162,.4361239459)(-1.205045838,-3.936123946)
\psline{->}(-.4633436855,-4.768328157)(.4633436855,-5.231671843)
\psline{->}(-.6671798822,-4.999230177)(.6671798822,-7.000769823)
\psline{->}(-.4633436855,-.2683281572)(.4633436855,-.7316718428)
\psline{->}(-.7949541621,-.5638760541)(.7949541621,-4.936123946)
\psline{->}(-1.,2.4)(-1.,.6)
\psline{->}(1.,1.4)(1.,-.4)
\rput(-.3,-6){$\scriptstyle\frac{2}{3}$}
\rput(0,-4.65){$\scriptstyle\frac{2}{3}$}
\rput(.8,-6.4){$\scriptstyle\frac{2}{5}$}
\rput(.8,-4.15){$\scriptstyle\frac{2}{3}$}
\rput(.8,-1.9){$\scriptstyle\frac{2}{5}$}
\rput(-2.2,-1.7){$\scriptstyle\frac{2}{3}$}
\rput(-.8,-2.2){$\scriptstyle\frac{2}{5}$}
\rput(-.8,1.5){$\scriptstyle\frac{2}{3}$}
\rput(-2.1,.25){$\scriptstyle\frac{2}{3}$}
\rput(-.1,-.75){$\scriptstyle\frac{2}{3}$}
\rput(.1,-2.3){$\scriptstyle\frac{2}{3}$}
\rput(1.2,.5){$\scriptstyle\frac{2}{3}$}
}
\endpspicture
}
\def\supminmodprocI{\pspicture(-6,-2.5)(6,2.5)
\rput(-3.5,0){
\pscircle[linecolor=\backgroundcolor,linewidth=1pt](0,0){2}
\psset{dotsize=5pt 0}
\psdots(0,2)(1.902113032,.618033989)(1.175570504,-1.618033989)(-1.175570504,-1.618033989)(-1.902113032,.618033989)
\psset{linecolor=\branecolor,linewidth=2pt}
\psline(-1.902113032,.618033989)(0,2)(1.902113032,.618033989)(1.175570504,-1.618033989)
\psline{->}(-1.902113032,.618033989)(-.6340376773,1.539344663)
\psline{->}(0,2)(1.268075355,1.078689326)
\psline{->}(1.902113032,.618033989)(1.417751347,-.8726779963)}
\rput(3.5,0){
\pscircle[linecolor=\backgroundcolor,linewidth=1pt](0,0){2}
\psset{dotsize=5pt 0}
\psdots(0,2)(1.902113032,.618033989)(1.175570504,-1.618033989)(-1.175570504,-1.618033989)(-1.902113032,.618033989)
\psset{linecolor=\branecolor,linewidth=2pt}
\psline(-1.902113032,.618033989)(1.175570504,-1.618033989)
\psline{->}(-1.902113032,.618033989)(.149675992,-.8726779963)
}
\psline[linewidth=1.5pt,linecolor=\flowcolor]{->}(-0.8,0)(0.8,0)
\endpspicture
}
\def\supminmodprocII{\pspicture(-6,-2.5)(6,2.5)
\rput(-3.5,0){
\pscircle[linecolor=\backgroundcolor,linewidth=1pt](0,0){2}
\psset{dotsize=5pt 0}
\psdots(0,2)(1.902113032,.618033989)(1.175570504,-1.618033989)(-1.175570504,-1.618033989)(-1.902113032,.618033989)
\psset{linecolor=\branecolor,linewidth=2pt}
\psline(0,2)(1.175570504,-1.618033989)
\psline(1.902113032,.618033989)(-1.175570504,-1.618033989)
\psline{->}(0,2)(.6465637772,.010081306)
\psline{->}(1.902113032,.618033989)(-.1496759920,-.8726779963)}
\rput(3.5,0){
\pscircle[linecolor=\backgroundcolor,linewidth=1pt](0,0){2}
\psset{dotsize=5pt 0}
\psdots(0,2)(1.902113032,.618033989)(1.175570504,-1.618033989)(-1.175570504,-1.618033989)(-1.902113032,.618033989)
\psset{linecolor=\branecolor,linewidth=2pt}
\psline(0,2)(-1.175570504,-1.618033989)
\psline(1.902113032,.618033989)(1.175570504,-1.618033989)
\psline{->}(0,2)(-.7837136693,-.4120226593)
\psline{->}(1.902113032,.618033989)(1.417751347,-.8726779963)
}
\psline[linewidth=1.5pt,linecolor=\flowcolor]{->}(-0.8,0)(0.8,0)
\endpspicture
}
\def\supminmodprocIII{\pspicture(-6,-2.5)(6,2.5)
\rput(-3.5,0){
\pscircle[linecolor=\backgroundcolor,linewidth=1pt](0,0){2}
\psset{dotsize=5pt 0}
\psdots(0,2)(1.902113032,.618033989)(1.175570504,-1.618033989)(-1.175570504,-1.618033989)(-1.902113032,.618033989)
\psset{linecolor=\branecolor,linewidth=2pt}
\psline(0,2)(1.175570504,-1.618033989)
\psline{->}(0,2)(.6465637772,.010081306)
\psline(1.902113032,.618033989)(-1.175570504,-1.618033989)
\psline{->}(1.902113032,.618033989)(.2093870872,-.6118033989)
\psline(1.175570504,-1.618033989)(-1.902113032,.618033989)
\psline{->}(1.175570504,-1.618033989)(-.8762185203,-.1273220037)}
\rput(3.5,0){
\pscircle[linecolor=\backgroundcolor,linewidth=1pt](0,0){2}
\psset{dotsize=5pt 0}
\psdots(0,2)(1.902113032,.618033989)(1.175570504,-1.618033989)(-1.175570504,-1.618033989)(-1.902113032,.618033989)
\psset{linecolor=\branecolor,linewidth=2pt}
\psline(0,2)(-1.902113032,.618033989)
\psline{->}(0,2)(-1.268075355,1.078689326)
\psline(1.902113032,.618033989)(-1.175570504,-1.618033989)
\psline{->}(1.902113032,.618033989)(-.1496759920,-.8726779963)
}
\psline[linewidth=1.5pt,linecolor=\flowcolor]{->}(-0.8,0)(0.8,0)
\endpspicture
}
\def\supminmodprocIV{\pspicture(-6,-2.5)(6,2.5)
\rput(-3.5,0){
\pscircle[linecolor=\backgroundcolor,linewidth=1pt](0,0){2}
\psset{dotsize=5pt 0}
\psdots(0,2)(1.902113032,.618033989)(1.175570504,-1.618033989)(-1.175570504,-1.618033989)(-1.902113032,.618033989)
\psset{linecolor=\branecolor,linewidth=2pt}
\psline(0,2)(1.175570504,-1.618033989)
\psline{->}(0,2)(.6465637772,.010081306)
\psline(1.902113032,.618033989)(-1.175570504,-1.618033989)
\psline{->}(1.902113032,.618033989)(.2093870872,-.6118033989)
\psline(1.175570504,-1.618033989)(-1.902113032,.618033989)
\psline{->}(1.175570504,-1.618033989)(-.5171554412,-.3881966011)
\psline(-1.175570504,-1.618033989)(0,2)
\psline{->}(-1.175570504,-1.618033989)(-.3918568347,.79398867)
}
\rput(3.5,0){
\pscircle[linecolor=\backgroundcolor,linewidth=1pt](0,0){2}
\psset{dotsize=5pt 0}
\psdots(0,2)(1.902113032,.618033989)(1.175570504,-1.618033989)(-1.175570504,-1.618033989)(-1.902113032,.618033989)
\psset{linecolor=\branecolor,linewidth=2pt}
\psline(1.902113032,.618033989)(-1.902113032,.618033989)
\psline{->}(1.902113032,.618033989)(-.6340376773,.618033989)
}
\psline[linewidth=1.5pt,linecolor=\flowcolor]{->}(-0.8,0)(0.8,0)
\endpspicture
}
\def\supminmodprocV{\pspicture(-6,-2.5)(6,2.5)
\rput(-3.5,0){
\pscircle[linecolor=\backgroundcolor,linewidth=1pt](0,0){2}
\psset{dotsize=5pt 0}
\psdots(0,2)(1.902113032,.618033989)(1.175570504,-1.618033989)(-1.175570504,-1.618033989)(-1.902113032,.618033989)
\psset{linecolor=\branecolor,linewidth=2pt}
\psline(-1.902113032,.618033989)(0,2)(1.902113032,.618033989)(1.175570504,-1.618033989)(-1.175570504,-1.618033989)
\psline{->}(-1.902113032,.618033989)(-.6340376773,1.539344663)
\psline{->}(0,2)(1.268075355,1.078689326)
\psline{->}(1.902113032,.618033989)(1.417751347,-.8726779963)
\psline{->}(1.175570504,-1.618033989)(-.3918568347,-1.618033989)}
\rput(3.5,0){
\pscircle[linecolor=\backgroundcolor,linewidth=1pt](0,0){2}
\psset{dotsize=5pt 0}
\psdots(0,2)(1.902113032,.618033989)(1.175570504,-1.618033989)(-1.175570504,-1.618033989)(-1.902113032,.618033989)
\psset{linecolor=\branecolor,linewidth=2pt}
\psline(-1.902113032,.618033989)(-1.175570504,-1.618033989)
\psline{->}(-1.902113032,.618033989)(-1.417751347,-.8726779963)
}
\psline[linewidth=1.5pt,linecolor=\flowcolor]{->}(-0.8,0)(0.8,0)
\endpspicture
}

\newcommand{\ZZ}{\mathbb{Z}}

\newcommand{\NN}{\mathbb{N}}

\newcommand{\Lie}[1]{{\mathfrak{#1}}}
\newcommand{\ALie}[1]{\widehat{\mathfrak{#1}}}
\newcommand{\mc}{\mathcal} 

\newcommand{\sspin}{S}
\newcommand{\lra}{\longrightarrow}

\newcommand{\simp}{\mc{J}}

\newcommand{\crep}[1]{{#1{}^{\raisebox{.3mm}{\scalebox{0.55}{$\scriptstyle
+$}}}}}
\newcommand{\creplarge}[1]{{#1{}^{\raisebox{.3mm}{\scalebox{0.55}{$\textstyle
+$}}}}}
\newcommand{\negsp}{\mspace{-1.2mu}}
\newcommand{\possp}{\mspace{1.2mu}}

\def\bL{\mc{B}} 
\def\Rep{{\text{Rep}}}
\def\tR{{\rm R}} 
\def\id{{\rm id}}

\def\Ad{{\rm Ad}}

\def\cH{{\cal H}} 
 
\def\cE{{\cal E}} 
\def\tA{{\rm A}}  
\def\cS{{\cal S}} 
\newcommand{\tr}{\mbox{tr}}

\begin{document}
\baselineskip=17pt
\title{\bf Organizing boundary RG flows}

\author {{\sc Stefan Fredenhagen} \\[2mm]
                  CPHT -- Ecole Polytechnique\\
                  F--91128 Palaiseau CEDEX, France}
\vskip.2cm

\date{January 28, 2003}
\begin{titlepage}      \maketitle       \thispagestyle{empty}

\vskip1cm
\begin{abstract}
\noindent 
We show how a large class of boundary RG flows in 
two-dimensional conformal field theories can be summarized in a single rule.
This rule is a generalization of the 
'absorption of the boundary spin'-principle of Affleck and Ludwig and 
applies to all theories which have a description as a coset model. 
We give a formulation for coset models with arbitrary modular 
invariant partition function and present evidence for the conjectured rule.
The second half of the
article contains an illustrated section of examples where the
rule is applied to unitary minimal models of the A- and D-series, in
particular the 3-state Potts model, and to parafermion theories. We
demonstrate how the rule can be used to compute brane charge groups in
the example of $N=2$ minimal models.
\end{abstract}
\vspace*{-15.9cm}
{\tt {{CPHT-RR-006-0103} \hfill hep-th/0301229}}
\bigskip\vfill
\noindent
{\small e-mail:}{\small\tt stefan@cpht.polytechnique.fr} 
\end{titlepage}
\section{Introduction}
The study of renormalization group (RG) flows in two-dimensional quantum
field theories is an important subject in condensed matter
physics and statistical mechanics, and it also plays a vital role in
string theory. In systems with boundaries or
defects, there are flows generated by boundary fields 
which only affect the boundary condition and leave the theory in the
bulk unchanged. In string theory, such flows describe the dynamics of
D-branes in a given closed string background. 

How do we find boundary RG flows? For a given boundary perturbation of
a boundary conformal field theory (BCFT), we have various tools at our
disposal. In some cases when we perturb by a field which is 
only `slightly' relevant, we can apply
perturbation theory~\cite{Affleck:1991tk}. If the perturbation is
integrable, we may use exact integral equation techniques (like the
Thermodynamic Bethe Ansatz). A method which can always be
employed is the Truncated Conformal Space Approach where we
truncate the Hilbert space to a finite-dimensional space and compute
RG flows numerically.

All these tools have helped to get a substantial knowledge about
boundary RG flows. To organize the informations, we need general,
model-independent principles. One such principle is the `g-conjecture'
of Affleck and Ludwig~\cite{Affleck:1991tk} which states that the
boundary entropy $g$ always decreases along a RG flow. Although very
important, the `g-conjecture' is not a constructive principle: it only
tells us which flows are \textit{possible} and which are not. 

In the case of WZNW models, we have a constructive principle at
hand, namely the `absorption of the boundary spin'-principle of
Affleck and Ludwig~\cite{Affleck:1991by}. This rule is easy to formulate and
describes a large class of flows. 

A generalization of this rule to fixed-point free coset models was proposed
in~\cite{Fredenhagen:2002qn}. The formulation there was for coset
models with a charge-conjugated modular invariant partition function
and boundary conditions of Cardy type. Here, we shall present a
formulation that is applicable for all maximally symmetric boundary
conditions in coset models with any modular invariant. Furthermore we
shall work out some arguments supporting the proposal, and employ the
rule in a number of examples.

The structure of the paper is as follows: We start with an introduction
to coset models and their maximally symmetric boundary conditions 
in section~2. Although this is essentially a review
of~\cite{Ishikawa:2001zu,Ishikawa:2002wx}, we hope to
clarify the role of the modular invariant when we relate coset
boundary conditions to those of WZNW models. In section 3, we
formulate the `absorption of boundary spin'-principle and its
generalization to coset models and discuss the relation to
perturbative calculations and the compatibility with the g-conjecture.
Through a number of examples, we present the rule at work in section~4. We
shall make extensive use of 
a geometric interpretation of the boundary conditions as
`branes' to visualize the RG flows. Whenever we are aware of results on
boundary RG flows in the specific examples, we compare them to the
predictions of the rule. At the end of section~4 the rule is used to
determine the charge group of branes in $N=2$ minimal models. 
In the appendix we collect the complete
results for the critical and tricritical Ising model as
well as the 3-state Potts model.

\section{Boundary Conditions in coset models}
The coset construction~\cite{Goddard:1985vk} allows to access a great 
variety of rational conformal field theories. 
Boundary conditions in these models have been investigated in the
past. Most of the work was concentrated on maximally symmetric 
boundary conditions, i.e.\ those
where the boundary conformal field theory admits the action of the
coset chiral algebra $\ALie{g}/\ALie{h}$. Untwisted boundary
conditions in coset models with charge-conjugated modular invariant
partition function are already covered by the seminal paper of Cardy on
boundary conditions in rational CFTs~\cite{Cardy:1989ir}. The
generalization to twisted boundary conditions and more general modular
invariants has been worked out
in~\cite{Ishikawa:2001zu,Ishikawa:2002wx}. Symmetry breaking boundary
conditions in coset models have been first considered
in~\cite{Quella:2002ns,Quella:2002fk} relying on previous work in WZNW 
models~\cite{Maldacena:2001ky,Quella:2002ct}. In the $\sigma $-model approach,
boundary conditions in gauged WZNW-models have been studied
in~\cite{Gawedzki:2001ye,Elitzur:2001qd,Kubota:2001ai}. Recently there
has been also some work on boundary conditions in asymmetric cosets~\cite{Walton:2002db,Sarkissian:2002nq,Quella:2002fk}.
\smallskip

We give a short introduction to coset models to set up our
notation. Subsequently, we discuss some general properties of maximally
symmetric boundary conditions. The section ends with a review on 
the construction
of boundary states from known boundary conditions in the product
theory with chiral algebra $\ALie{g}\oplus \ALie{h}$.

\subsection{Coset construction}
From now on let $\Lie{h} \subset \Lie{g}$ denote some simple 
subalgebra of the simple Lie algebra $\Lie{g}$ (the generalization to
semi-simple Lie algebras is straightforward). We want to study the associated 
$\ALie{g}/\ALie{h}$ coset model. A more precise formulation of this theory 
requires a bit of preparation (more details can be found e.g.\ 
in \cite{FrancescoCFT}). 

Induced from the embedding of $\Lie{h} $ in $\Lie{g}$, there is an embedding of
the affine Lie algebra $\ALie{h}_{k'}$ into $\ALie{g}_{k}$. The level
$k'$ is related to $k$ by the embedding index $x_{e}$, $k'=kx_{e}$.
We shall label the sectors $\cH_\Lie{h}^{l'}$ 
of the affine Lie algebra $\ALie{h}_{k'}$ with labels $l' \in 
\Rep (\ALie{h}_{k'})$. Note that the sectors of the numerator theory 
carry an action of the denominator algebra $\ALie{h}_{k'} \subset 
\ALie{g}_k$ and under this action each sector $\cH_\Lie{g}^{l}$ 
decomposes according to 
\begin{equation}\label{numeratorexpansion}
 \cH_\Lie{g}^{l} \ = \ \bigoplus_{l'} \ \cH^{(l,
     l')} \otimes \cH_\Lie{h}^{l'}\ \ . 
\end{equation}
Here we have introduced the infinite dimensional spaces 
$\cH^{(l,l')}$ which we want to interpret as 
sectors of the coset chiral algebra. The latter is usually 
hard to describe explicitly, but at least it is known to 
contain a Virasoro field with modes
\begin{equation}\label{eq:cosetvirasoro}
 L_n \ = \ L^\Lie{g}_n - L^\Lie{h}_n \ \ . 
\end{equation}
One may easily check that they obey the usual exchange 
relations of the Virasoro algebra with central charge given 
by $c = c^\Lie{g} -c^\Lie{h}$. 
\smallskip

Note that some of the spaces $\cH^{(l,l')}$ may be trivial 
simply because a given sector $\cH_\Lie{h}^{l'}$ of the 
denominator theory may not appear as a subsector in a 
given $\cH_\Lie{g}^{l}$. This allows to introduce the set
\[
\cE \ = \ \{ \, (l,l') \in \Rep (\ALie{g}_{k}) \times 
 \Rep (\ALie{h}_{k'})  \, | \, \cH^{(l,l')} \neq 0\, 
    \} \ \ .
\]
Furthermore, some of the coset spaces labeled by
different pairs $(l,l')$ and $(m,m')$ correspond to
the same sector of the coset theory. Therefore we label coset
sectors by equivalence classes $[l,l']$ of pairs.

There is an elegant formalism to describe these selection and
identification rules which is
applicable in almost all coset models\footnote{The known exceptions
all appear at low levels of the involved affine Lie algebras, see e.g.\
the Maverick cosets~\cite{Dunbar:1993hr}}. It involves the so-called
identification group $\mc{G}_{\id }$ which contains pairs $(\simp
,\simp ')$ of simple currents. It is a subgroup of the direct
product of the simple current groups of $\ALie{g}_{k}$ and
$\ALie{h}_{k'}$. A simple current $\simp $ 
of $\ALie{g}_{k}$ is an element in $\Rep(\ALie{g}_{k})$ which has the
property that the fusion product of $\simp $
with any other representation $l$ contains exactly one sector $m=:\simp
\negsp l\in \Rep (\ALie{g}_{k})$,
\[
{N_{\simp l}}^{m}\ =\ \delta_{m,\simp\negsp  l}\ \ .
\]
To formulate the selection rules in coset models, we introduce
the monodromy charge $Q_{\simp } (l)$ of $l$ with respect to $\simp $
in terms of conformal weights,
\[
Q_{\simp } (l)\ =\ h_{\simp }+h_{l}-h_{\simp\negsp l}\ \ \mod \ \ZZ \ \ .
\]
The monodromy charge appears when a simple current $\simp $ acts on
the modular S-matrix, 
\[
S_{\simp\negsp l\, m}\ =\ e^{2\pi iQ_{\simp } (m)}\, S_{l\,m}\ \ .
\]

We are now prepared to formulate selection and identification rules in
terms of the identification group $\mc{G}_{\id }$ of simple currents:
\begin{itemize}
\item A pair $(l,l')$ is allowed, i.e.\ $(l,l')\in \cE $, if
$Q_{\simp } (l)=Q_{\simp '} (l')$ for all $(\simp ,\simp ')\in
\mc{G}_{\id }$
\item Two pairs $(l,l')$ and $(\simp\negsp l,\simp'\negsp l')$ 
label the same sector, i.e.\ 
\[
\cH ^{(l,l')}\ \cong \ \cH ^{(\simp \negsp  l,\simp'\negsp  l')}\ \ .
\]
\end{itemize}
At this point we want to make one assumption, 
namely that all the equivalence classes we find in $\cE$ 
contain the same number $N_{0}=|\mc{G}_{\id }|$ of elements, in other
words, $\mc{G}_{\id }$ acts fixed-point free. 
This holds true for many 
important examples and it guarantees that the sectors of the 
coset theory are simply labeled by the equivalence classes\footnote{For more 
general cases, there are 
further sectors that cannot be constructed within the sectors 
of the numerator theory.}, 
i.e.\ $\Rep (\ALie{g}/\ALie{h}) = \cE/\mc{G}_{\id }$. It is then 
also easy to spell out explicit formulas for the fusion rules and the S-matrix 
of the coset model. These are given by
\begin{eqnarray} 
N_{[j,j'] [k,k']}^{\quad \quad \quad [l,l']} & = & 
 \sum_{(m ,m ')\sim (l , l ')}   
N^{\Lie{g}}_{j\, k}{}^{m}\, N^{\Lie{h}}_{j'\,k'}{}^{m'} \ \ , 
\label{eq:cosN} \\[2mm] 
S_{[l,l'][m,m']} & = & 
 N_{0}\, S^\Lie{g}_{l\,m} \ \bar S^{\Lie{h}}_{l'\, m'} \ \ 
\label{eq:cosS} \end{eqnarray} 
where the bar over the second S-matrix denotes complex 
conjugation. 

\subsection{Maximally symmetric boundary conditions}
Let us turn now to coset models with a boundary and summarize
some general properties of maximally symmetric boundary conditions 
to be prepared for the concrete analysis in section~2.3. 

We want to impose
conditions along the boundary gluing left moving and right moving
fields together with a suitable automorphism $\omega $ 
of the coset chiral algebra. The corresponding set of elementary
boundary conditions is denoted by $\bL^{\omega }_{\Lie{g}/\Lie{h}}$.

Because of the specific gluing conditions, the annulus partition
function involving the boundary conditions $\alpha ,\beta \in \bL
^{\omega }_{\Lie{g}/\Lie{h}}$ decomposes into coset characters,
\[
{Z_{\alpha }}^{\beta } (q)\ =\ \sum_{[l,l']}\ n_{[l,l']\, \alpha
}{}^{\beta } \chi ^{(l,l')} (q)
\]
with non-negative integers $n_{[l,l']\, \alpha
}{}^{\beta }$. For a complete set of boundary conditions $\bL ^{\omega
}_{\Lie{g}/\Lie{h}}$, these numbers
are known to form a representation of the fusion algebra
\cite{Behrend:1998fd,Behrend:1999bn},
\[
\sum_{\beta }\ n_{[l,l']\, \alpha }{}^{\beta } n_{[j,j']\, \beta
}{}^{\gamma } \ =\ \sum_{[k,k']}\  N_{[l,l'][j,j']}{}^{[k,k']} 
 n_{[k,k']\, \alpha}{}^{\gamma }\ \ .
\]
The integers $n$ have the further properties 
\[
\makebox[1cm][r]{$\displaystyle n_{[0,0]\alpha }{}^{\beta }$}\ =\
\makebox[1cm][l]{$\displaystyle\delta _{\alpha \beta }$}
\]
and
\[
\makebox[1cm][r]{$\displaystyle n_{[l,l']\alpha }{}^{\beta }$}\ =\
\makebox[1cm][l]{$\displaystyle n_{[\crep{l},\crep{l'}]\beta
}{}^{\alpha }$} 
\]
where $\creplarge{l}$ labels the representation conjugate to $l$.

\subsection{Boundary conditions from WZNW models}
In the last subsection we have been rather general. 
It is possible to relate the analysis of boundary conditions in coset
models to the investigation of boundary conditions in the product
theory with chiral algebra $\ALie{g}\oplus \ALie{h}$
\cite{Fredenhagen:2001kw,Ishikawa:2001zu,Ishikawa:2002wx}. Before we enter
the detailed description, let us sketch our general procedure: 
we specify a modular-invariant partition function of
the coset model and from that we construct a partition function for the
product theory. In the resulting theory we impose gluing conditions involving a
gluing automorphism $\omega ^{\Lie{g}}\times (\omega ^{\Lie{h}})^{-1}$ where we assume
that $\omega ^{\Lie{g}}$ restricts to $\Lie{h}$ and $\omega
^{\Lie{h}}=\omega ^{\Lie{g}}|_{\Lie{h}}$. The corresponding boundary
conditions of the product theory can be projected to boundary
conditions in the coset model by certain selection rules.

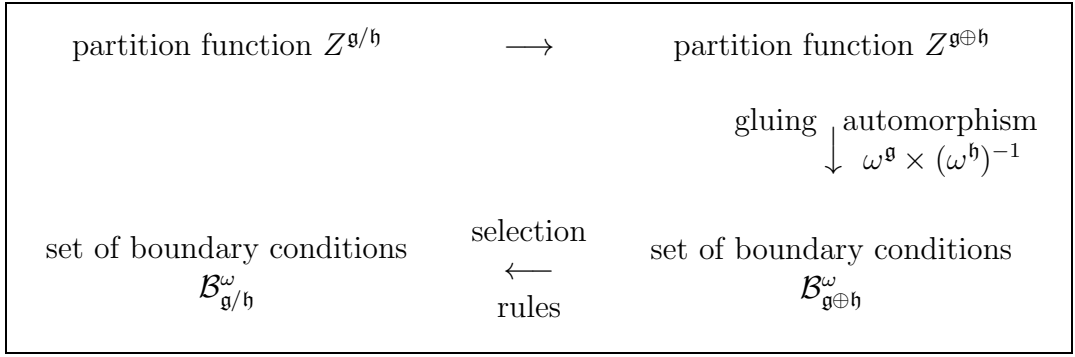
\begin{center}
\framebox{\parbox[c][2.1\height][c]{.93\textwidth}{
\begin{tabular}{ccc}
partition function $Z^{\Lie{g}/\Lie{h}}$ & $\longrightarrow$ &
partition function $Z^{\Lie{g}\oplus \Lie{h}}$\\[5mm]
 & & \rput{-90}{$\longrightarrow$} 
\makebox[0cm][l]{\begin{tabular}{c}
\hspace*{-1.5cm}gluing \hspace*{2mm}automorphism\\ $\omega ^{\Lie{g}}\times (\omega
^{\Lie{h}})^{-1}$\end{tabular}}
\\[7mm] \begin{tabular}{c}set of boundary conditions\\ $\bL ^{\omega
}_{\Lie{g}/\Lie{h}}$ \end{tabular} & \begin{tabular}{c}selection \\
$\longleftarrow $\\  rules\end{tabular} &
\begin{tabular}{c}set of boundary conditions\\ $\bL ^{\omega
}_{\Lie{g}\oplus \Lie{h}}$ \end{tabular}
\end{tabular}}}
\end{center}

Let us become more specific.
We start with a coset model with the partition function
\[
Z^{\Lie{g}/\Lie{h}} (q,\bar{q})\ =\ \sum_{[l,l'],[m ,m']}
Z_{[l ,l'],[m ,m']}\ \chi ^{[l ,l']}
(q)\, \chi^{[m ,m']} (\bar{q}) \ \ 
\]
with some non-negative integers $Z_{[l ,l'],[m ,m']}$.
To this model we associate a product theory $\Lie{g}\oplus \Lie{h}$
with partition function (following~\cite{Gannon:1995km})
\[
Z^{\Lie{g}\oplus \Lie{h}} (q,\bar{q})\ =\ \sum_{l,l', m ,m'}
Z_{[l ,l'],[m ,m']}\ \chi ^{l} (q) \chi ^{m'} (q)
\, \chi^{m}(\bar{q}) \chi ^{l'} (\bar{q}) \ \ .
\]
Note the exchange of the $\Lie{h}$-labels $l'$ and $m'$ which is
necessary to guarantee modular invariance of the product theory.

In this theory we want to analyze maximally symmetric boundary
conditions. To this end, we glue the left- and right-moving currents
$J (z),\bar{J} (\bar{z})$ of the $\ALie{g}$ and $\ALie{h}$ theory 
with a gluing automorphism
$\Omega$ indcued from $\omega ^{\Lie{g}}\times (\omega ^{\Lie{h}})^{-1}$ along
the boundary, 
\[
J (z)\ =\ \Omega \bigl( \bar{J} \bigr) (\bar{z})\ \ \ \  \mbox{ for } z=\bar{z}\ \ .
\]
Assume now that we have solved the problem of finding all maximally symmetric
boundary conditions in the theory, i.e.\  we have a set of boundary
conditions $\alpha \in \bL^{\omega }_{\Lie{g}\oplus \Lie{h}}$
specified by the boundary couplings $\psi _{\alpha }^{(l,l';\lambda
)}$. The corresponding boundary state is 
\[
|\alpha \rangle\ =\ \sum_{(l,l';\lambda )} \frac{\psi _{\alpha
}^{(l,l';\lambda )}}{\sqrt{S_{l0}S_{l'0}}}\, | (l,l';\lambda
)\rangle\!\rangle\ \ .
\]
Here, $(l,l';\lambda )$ labels an Ishibashi state in the sector
$(l,l')$ and $\lambda $ is an additional multiplicity index in the range 
\[
\lambda =1,\dots, Z_{[l,\possp \omega ^{-1} (\crep{l'})][\omega
(\crep{l}),\possp l']}\
\ .
\]

At this point we want to make an important assumption. We assume that
we can find a basis of Ishibashi states s.t.\ the action of a simple
current on the boundary couplings $\psi $ is given by a pure phase
factor which only depends on $\alpha $,
\[
\psi_{\alpha } ^{(\simp \negsp  l ,\possp \omega (\crep{\simp
'})\negsp l'; \lambda )}\ =\ e^{2\pi i Q_{(\simp
,\omega (\crep{\simp '}))} (\alpha )}\ \psi_{\alpha } ^{(l ,
l';\lambda )} \quad \mbox{for } (\simp ,\simp ') \in \mc{G}_{\id } \ \ .
\]
This is certainly true in all examples that we considered, but it is
unclear whether this assumption holds in general (this problem has
already been mentioned in~\cite{Ishikawa:2001zu,Ishikawa:2002wx}).

We are now prepared to write down a set of boundary conditions for the
coset model. For any $\alpha \in \bL^{\omega } _{\Lie{g}\oplus \Lie{h}}$ 
satisfying the selection rule
\begin{equation}\label{selrules}
Q_{(\simp ,\possp \omega (\crep{\simp'} ))} (\alpha )=0 \quad \text{ for all }
(\simp ,\simp ')\in \mc{G}_{\id }\ \ ,
\end{equation}
we define a boundary condition in the coset model
which we also label by $\alpha $,
\begin{equation}\label{cosetbcdef}
\psi _{\alpha }^{([l,l'];\lambda) }\ :=\ \sqrt{N_{0}}\ \psi _{\alpha
}^{(l,\possp \omega
(\crep{l'});\lambda )}\ \ .
\end{equation}
The Ishibashi states of the coset model are labeled by equivalence
classes of pairs $[l,l']$ together with a multiplicity index $\lambda
$ running from 1 to $Z_{[l,l'],[\omega (\crep{l}),\omega
(\crep{l'})]}$. 
Note that the multiplicity of the coset Ishibashi state $[l,l']$
and the product Ishibashi state $( l,\omega (\crep{l'}))$ coincide
so that we can use the same label $\lambda $ on both sides of eq.\
\eqref{cosetbcdef}.

It is straightforward to verify that the $\psi _{\alpha
}^{([l,l'];\lambda) }$ fulfill the completeness conditions
\begin{align}
\sum_{\alpha } \psi _{\alpha }^{([l,l'];\lambda) }\overline{\psi}{} _{\alpha
}^{([m,m'];\mu) }\ & =\ \delta _{[l,l'],[m,m']}\delta _{\lambda
,\mu }\\ 
\intertext{and}
\sum_{([l,l'];\lambda)} \psi _{\alpha }^{([l,l'];\lambda) }\overline{\psi}{}
_{\beta }^{([l,l'];\lambda) }\ & =\ \delta _{\alpha ,\beta }\ \ .
\end{align}
Furthermore, Cardy's condition which says that the annulus coefficients 
\[
n^{\Lie{g}/\Lie{h}}_{[l,l']\alpha }{}^{\beta }\ =\
\sum_{([j,j'];\lambda )} \frac{\psi _{\alpha }^{([j,j'];\lambda
)}\overline{\psi}{}
_{\beta }^{([j,j'];\lambda
)} S^{\Lie{g}/\Lie{h}}_{[l,l'][j,j']}}{S^{\Lie{g}/\Lie{h}}_{[0,0][j,j']}}
\ \ 
\]
are non-negative integers, is satisfied, and 
\begin{equation}\label{anncoeff}
n_{[l,l']\alpha }^{\Lie{g}/\Lie{h}}{}^{\beta }\ =\ n_{(l,\omega (l'))\alpha
}^{\Lie{g}\oplus \Lie{h}}{}^{\beta } \ \ .
\end{equation}

\subsection{The factorizing case}
In this section we want to deal with the situation of a `factorizing'
modular invariant partition function in the coset theory. We should
state more clearly what we mean by `factorizing', namely 
that the associated modular invariant of the product theory is a
$\tilde{\mc{G}}_{\id }$ simple current orbifold of a direct product of a
$\ALie{g}$ and a $\ALie{h}$ modular invariant,
\[
Z^{\Lie{g}\oplus \Lie{h}}\ =\ \sum_{(\simp ,\simp ')\in \tilde{\mc{G}}_{\id }}
\hspace*{-1mm}\sum_{\substack{l,l',m,m'\\
Q_{\simp } (l)+Q_{\simp '} (l')=0}} \hspace*{-4mm}
Z^{\Lie{g}}_{l\, \simp\! m}\,
Z^{\Lie{h}}_{l'\, \simp '\negsp m'}\ \chi^{l} (q)\chi^{l'} (q)\ \chi^{m}
(\bar{q})\chi^{m'} (\bar{q})\ \  
\]
where $\tilde{\mc{G}}_{\id }=\{(\simp ,\simp ') | (\simp ,\simp
'^{-1})\in \mc{G}_{\id } \}$.
The boundary conditions $\alpha \in \bL ^{\omega }_{\Lie{g}\oplus
\Lie{h}}$ in the simple current orbifold can be obtained
from pairs $(L ,L')$ of boundary conditions  $L \in \bL ^{\omega
}_{\Lie{g}}$ 
and $L '\in \bL ^{\omega }_{\Lie{h}}$ of the $\ALie{g}$ and $\ALie{h}$
theory, respectively. These pairs are subject to identification rules,
and fixed-points can occur (even if $\mc{G}_{\id }$ acts fixed-point
free on the sectors). These orbifold fixed-points can be easily 
resolved\footnote{In string theory these resolved boundary conditions
are called fractional branes.}. 
We shall label the boundary conditions by
equivalence classes of pairs $\alpha =[L ,L '] $, always remembering 
the possible orbifold fixed-point resolution.

When we want to obtain boundary conditions in the coset model along
the lines of section~2.3, we only
have to impose in addition the selection rules~\eqref{selrules} 
on the boundary conditions $\alpha = [L ,L ']$. 

We arrive at the final conclusion that -- in the factorizing case --
boundary conditions of the coset model are obtained from pairs of
boundary conditions of numerator and denominator theory by suitable
identification and selection rules. This result has been formulated
first in~\cite{Ishikawa:2001zu}. There, the authors took a direct
way not involving boundary conditions in the simple current orbifold.
In practice this can simplify things: it is not always
necessary to do the fixed-point resolution in the orbifold step,
because it may happen that many of the resolved boundary
conditions do not survive the selection rules. 
Still we think that the detour via the simple current orbifold has
conceptual advantages. Firstly, it shifts all problems with
fixed-point resolution to the orbifold step. Secondly, it fits in the
general framework of section~2.3 which is also applicable in the
non-factorizing case.

In all our examples in section~4, we
shall encounter the factorizing case. The formulation of the 
rule for boundary RG flows in section~3, however, is more general and
can be also used in the non-factorizing case.


\section{RG flows: a simple rule}
\subsection{Generalized Affleck-Ludwig rule}
To motivate the rule for boundary RG flows in coset models, we shall
first review shortly the original proposal of Affleck and Ludwig for
the absorption of the boundary spin in the Kondo model.

The Kondo model is designed to understand the effect of 
magnetic impurities on the low temperature conductivity
of a conductor. Usually a decreasing temperature will result in an
increasing conductivity, because the scattering with phonons is
reduced (Matthiesen's rule). In some cases,
however, when magnetic impurities are present, the conductivity reaches a
maximum and starts to decrease again. This phenomenon is explained by
the coupling of the electrons to the magnetic impurities. The electrons
tend to screen the impurity, and this coupling increases when
temperatures become low.

Let us say that the conductor has electrons in 
$k$ conduction bands. We can build several currents from 
the basic fermionic fields: the charge current, the flavor current, and the 
spin current $\vec{J}(y)$. 
The latter gives rise to a $\widehat{su}(2)_k$ 
current algebra. The coordinate $y$ measures the radial 
distance from a spin~$\sspin$ impurity at $y=0$ to which 
the spin current couples\footnote{We only consider the case of a
single isolated impurity.}. This coupling is 
\begin{equation} \label{Kondo} 
    H_{\rm pert} \ = \ \lambda \ \tR_\alpha  J^\alpha (0) \ \ . 
\end{equation}
where $\tR_{\alpha }\ (\alpha =1,2,3)$ 
is a $2\sspin +1$ dimensional irreducible
representation of $su (2)$, $\lambda $ is the coupling constant.

The operator $H_{\rm pert}$ acts on the tensor product 
$V^{\sspin } \otimes \cH$ of the $2 \sspin  + 1$ -dimensional 
quantum mechanical state space of our impurity with the Hilbert 
space $\cH$ for the unperturbed theory described by a Hamiltonian
$H_{0}$. 

When the boundary spin is large ($2 \sspin  > k$),
the low temperature fixed point 
of the Kondo model appears only at infinite values
of $\lambda$ (`under-screening'). On the other hand, the fixed point 
is reached at a finite value $\lambda = \lambda^*$ of 
the renormalized coupling constant $\lambda$ if $ 2 \sspin  
\leq k$ (exact- or over-screening resp.). In the latter 
case, the fixed points are described by non-trivial
(interacting) conformal field theories. Affleck and 
Ludwig \cite{Affleck:1992ng,Affleck:1991by} found an
elegant rule to determine these strong-coupling fixed-points. 
The spectrum at the fixed-point is given by
\begin{equation} \label{charpert}  
  \tr_{V^{\sspin } \otimes \cH^{l}}\left( q^{H_0 + 
  H_{\rm pert}}\right)_{\lambda = \lambda^*}^{\rm ren} 
  = \sum_j {N_{\sspin  j}}^{l} \chi^{j}(q)\ \ .  
\end{equation} 
Here, $H_0 = L_0 + c/24$ is the unperturbed Hamiltonian, and
the superscript $\ ^{\rm ren}$ stands for `renormalized'.
By $\sspin $ we label a dominant highest-weight representation of
$\widehat{su} (2)$.
$V^{\sspin }$ denotes the corresponding module of the 
finite-dimensional Lie algebra $su(2)$, and $\cH ^{l}$ is an
irreducible sector of the $\widehat{su} (2)_{k}$-theory. 
The formula \eqref{charpert} is the
content of the `absorption of boundary spin'-principle by Affleck and
Ludwig~\cite{Affleck:1992ng,Affleck:1991by}.
\smallskip

It is straightforward to generalize these considerations 
to an arbitrary simple Lie algebra $\Lie{g}$. The space $\cH^{l}$ 
can be any of the $\ALie{g}_k$-irreducible subspaces in the 
physical state space $\cH$ of the theory. Formula \eqref{charpert} 
means that our perturbation with some irreducible representation 
$\sspin $ interpolates continuously between a building block 
$\dim(V^{\sspin }) \chi^{l}(q)$ of the partition function of the 
UV-fixed point (i.e.\ $\lambda = 0$) and the sum of characters 
on the right hand side of the  previous formula, 
\begin{equation} \label{abs} \dim (V^{\sspin }) \   
\chi^{l} (q)\ \longrightarrow\ \sum_j {N_{\sspin  j}}^{l} \chi^{j}(q)\ \ .
\end{equation}

In~\cite{Fredenhagen:2002qn} it was proposed to generalize the
`absorption of boundary spin'-principle to coset models. The suggested
rule is
\begin{equation} \label{abscoset}
\sum_{\sspin',l'}\ b_{\crep{\sspin}\sspin'}\ {N_{\sspin'l'}}^{j'}\
\chi^{(l,l')} (q)\ \longrightarrow \ \sum_{j}\ {N_{\sspin j}}^{l}\
\chi ^{(j,j')} (q)\ \ . 
\end{equation}
Here, $\sspin,l $ and $j'$ label dominant highest-weight representations of
$\ALie{g}$ and $\ALie{h}$, respectively. The 
coefficients $b_{\sspin\sspin'}$ are the branching
coefficients describing the decomposition of $V^{S}$, the corresponding
representation of the finite Lie algebra $\Lie{g}$, into
representations $V^{S'}$ of $\Lie{h}$,
\begin{equation}\label{repdecomposition}
V^{S}\ =\ \bigoplus b_{SS'}\ V^{S'}\ \ .
\end{equation}
The embedding of
affine Lie algebras $\ALie{h}\subset\ALie{g}$ guarantees that these 
representations can again be identified with 
highest-weight representations $\cH^{S'}$ of $\ALie{h}$. 

The flows (\ref{abscoset}) 
are generated by fields coming from the coset sectors 
\begin{equation} \label{list}  
\mathcal{H} ^{(0,l')}\ \ , \ \ \mbox{ where } \ \ 
   V^{l'} \subset V^{\theta} |_{\mathfrak{h}} \ \ \  .   
\end{equation} 
Here, $\theta$ labels the integrable highest-weight 
representation which is built from the 
adjoint representation of the Lie algebra $\mathfrak{g}$. 
The adjoint representation $l' = \theta'$ of $\Lie{h}$ can be omitted from
the list (\ref{list}) if it occurs only once in the decomposition of
$\theta $.
\smallskip

To see that \eqref{abscoset} is really a generalization of
\eqref{abs}, we should recover the flows \eqref{abs} when specializing
to the trivial subgroup $\{e \}$ of $\Lie{g}$. The primed label can
then be omitted and the branching coefficient is just the dimension of
the representation space $V^\sspin $.

\subsection{Simple rule for boundary RG flows in coset models}

From the stated rule~\eqref{abscoset} formulated in terms of 
characters we can infer
a rule for a flow between superpositions of boundary conditions. Before 
we discuss how this is done, let us present the result.

Choose a representation $S\in \Rep (\ALie{g})$ and a boundary
condition $\alpha \in \bL
^{\omega }_{\Lie{g}\oplus \Lie{h}}$ s.t.\ 
\[
Q_{(\simp ,\omega (\crep{\simp '} ))} (\alpha )+Q_{(\simp ,\omega (\crep{\simp'} ))} (S,0)\ =\ 0
\quad \text{for all } (\simp ,\simp ')\in \mc{G}_{\id }\ \ .
\]
Then there will be a RG flow between the following coset boundary 
configurations $X$ and $Y$,
\begin{equation}\label{bcflow}
X\ :=\ (0, \creplarge{S}|_{\Lie{h}}) \, \hat{\times }\,  \alpha \ \longrightarrow \ 
(S,0)\, \hat{\times }\,  \alpha  \ =:\ Y\ \ .
\end{equation}
Here, we introduced the shorthand notation $\hat{\times }$ to define a
superposition of the form
\[
(l,l')\, \hat{\times }\, \alpha \ := \ \bigoplus_{\beta }\ 
n^{\Lie{g}\oplus \Lie{h}}_{(l,l')\alpha }{}^{\beta }\ (\beta ) \ \ .
\]
The label $(0,\creplarge{S}|_{\Lie{h}})$ has to be understood as
\[
(0,\creplarge{S}|_{\Lie{h}})\hat{\times }\dots \ :=\ \bigoplus
b_{\crep{S}S'}\ (0,S')\hat
{\times }\dots \ \ 
\]
where $b_{\crep{S}S'}$ denote the finite branching coefficients defined
in~\eqref{repdecomposition}.
The flows~\eqref{bcflow}  
are generated by fields from the coset sectors~\eqref{list}.  
\smallskip

To derive~\eqref{bcflow} from~\eqref{abscoset}, we introduce an
arbitrary `spectator'
boundary condition $\beta $. The annulus partition function
$Z_{X}^{\ \beta } (q)$ can be decomposed into combinations of characters
that appear on the left side of~\eqref{abscoset} just using the fact
that the annulus coefficients form a representation of the fusion
algebra. 
The expression that
we obtain from applying~\eqref{abscoset} can then be rewritten as
$Z_{Y}^{\ \beta }$, i.e.\ we find the result
\[
Z_{X}^{\ \beta }\ \longrightarrow \ Z_{Y}^{\ \beta }
\]
for arbitrary boundary conditions $\beta $.
\medskip 

In the remainder of this section we want to give two arguments to
support our claim. First, we want to relate the rule to results
from a perturbative analysis in the limit when some levels are
large. Then we shall present evidence
that the rule is compatible with the g-conjecture of Affleck and
Ludwig.
\medskip

For a general coset theory with semi-simple numerator and denominator
there occur different levels $k_{r}$ for the simple constituents of
the numerator which then determine the levels in the denominator. 
Assume that we take some of the levels to very large values
of the order of a common scale $k\gg 1$. In the limit $k\to \infty $,
there are many coset fields whose conformal weight approaches $h=1$
(the difference to 1 being of the order $1/k$). The RG flows induced
by such fields can be studied by perturbative techniques. One
way is to use the method of effective actions. Here, the couplings of
the boundary fields are combined into matrices $\boldsymbol{\tA} $ which are
interpreted as fields in an effective theory determined by an action
$\cS [\boldsymbol{\tA} ]$. The equations of motion for $\boldsymbol{\tA} $ are precisely the
fixed-point equations $\beta =0$. 
In \cite{Fredenhagen:2001nc,Fredenhagen:2001kw} the effective action
for untwisted boundary conditions in coset models has been constructed to
leading order in $1/k$ building upon earlier works in 
WZNW models~\cite{Alekseev:2000fd}. The generalization to twisted
boundary conditions has been
worked out
in~\cite{Fredenhagen:thesis} using results of~\cite{Alekseev:2002rj}. 

A special class of solutions for all, untwisted and twisted, boundary
conditions~\cite{Fredenhagen:thesis} has precisely the form~\eqref{bcflow},
\[
(0, \creplarge{S}|_{\Lie{h}}) \, \hat{\times }\,  \alpha \ \longrightarrow \ 
(S,0)\, \hat{\times }\,  \alpha\ \ ,
\]
but here we have to restrict $S$ to representations s.t.\ the
conformal weight $h_{S}^{\Lie{g}}$ in the $\ALie{g}$-theory is of
order $1/k$. The rule~\eqref{bcflow} thus extrapolates the
perturbative results to arbitrary values of the levels.
\medskip

The g-conjecture of Affleck and Ludwig states that the boundary
entropy $g$ decreases along a boundary RG flow $X\longrightarrow Y$,
\[
g_{X}\ >\ g_{Y}\ \ .
\]
The boundary entropy $g_{X}$ for a superposition $X=\bigoplus_{\alpha
}X_{\alpha }\, \alpha $ (with $X_{\alpha } \in \NN_{0}$) of boundary
conditions $\alpha $ occurring with multiplicity $X_{\alpha }$ is defined as
the sum of the g-factors of the single boundary conditions,
\[
g_{X}\ =\ \sum_{\alpha }X_{\alpha }\, g_{\alpha }\ =\ \sum_{\alpha
}X_{\alpha }\, \frac{\psi _{\alpha}^{0}}{\sqrt{S_{00}}}\ \ .
\]
The ratio $g_{X}/g_{Y}$ for the conjectured flow~\eqref{bcflow} is
given by
\begin{align*}
\frac{g_{X}}{g_{Y}}&\ =\ \frac{\sum_{S'\beta
}b_{\crep{S}S'}\ n^{\Lie{g}\oplus \Lie{h}}_{(0,S')\alpha }{}^{\beta } \,
g_{\beta }}{\sum_{\gamma }n^{\Lie{g}\oplus \Lie{h}}_{(S,0)\alpha
}{}^{\gamma } \,g_{\gamma  } }\\
& \ =\ \frac{\sum_{S'\beta
}b_{\crep{S}S'}\ n^{\Lie{g}\oplus \Lie{h}}_{(0,S')\alpha }{}^{\beta } \,
\psi _{\beta }^{0}}{\sum_{\gamma }n^{\Lie{g}\oplus \Lie{h}}_{(S,0)\alpha
}{}^{\gamma } \,\psi _{\gamma  }^{0} }
\ \ .
\end{align*}
We simplify this expression by using the fact that the vector $(\psi
^{0})_{\beta }$ is an eigenvector of the matrix $(n_{(l,l')})_{\alpha
}{}^{\beta }$ with eigenvalue $S_{(l,l')0}/S_{00}$ and obtain a result
which does only depend on $S$,
\[
\frac{g_{X}}{g_{Y}}\ =\ \frac{\sum_{S'}b_{SS'}\,
S^{\Lie{g}}_{00}S^{\Lie{h}}_{S'0}}{S^{\Lie{g}}_{S0}S^{\Lie{h}}_{00}}\
\ .
\]
Hence, if our conjectured rule~\eqref{bcflow} and the g-conjecture 
are correct, we obtain the following inequality for quantum
dimensions of $\ALie{g}$ and $\ALie{h}$ ($S\not= 0 $),
\begin{equation}\label{inequality}
\sum_{S'}b_{SS'}\frac{S^{\Lie{h}}_{S'0}}{S^{\Lie{h}}_{00}}\
> \ \frac{S^{\Lie{g}}_{S0}}{S^{\Lie{g}}_{00}}\ \ .
\end{equation}
This inequality can be used to test our proposal.

For diagonal cosets $\ALie{g}_{k}\oplus
\ALie{g}_{l}/\ALie{g}_{k+l}$ the inequality is satisfied. This follows
from the fact that the quantum dimension
$S^{\Lie{g}_{k}}_{S0}/S^{\Lie{g}_{k}}_{00}$ of a fixed
representation $S$ is a monotonically increasing function of the level~$k$.
Unfortunately, for general coset models 
we have not found a proof yet. However, numerical checks have been performed in
a large number of coset models, all in accordance with the conjecture. 
Furthermore, when we take some levels to be large, we can confirm the 
inequality in a perturbative calculation (see appendix~A). 
\medskip

This ends our discussion of the general properties of the rule~\eqref{bcflow}.  
We have given some evidence by showing that the rule is consistent
with the perturbative results and, although not completely proven,
with the g-conjecture. The next section will provide more evidence
coming from specific examples where the rule is able to reproduce
a number of known flows.

\section{Examples}
In this section we shall present the rule~\eqref{bcflow} at work in a
number of examples. In all these examples we shall first give the
field content of the theory (coset sectors) and the boundary
conditions by specifying identification and selection rules. Then we
shall formulate the annulus coefficients for the coset boundary
theories. The boundary conditions for the associated product theory
are obtained by forgetting the selection rules on them, and the
corresponding annulus coefficients are related to the ones from the coset
model by \eqref{anncoeff}.
We shall introduce a pictorial
representation of the boundary conditions as branes in some target
space. This is followed by an application of the rule~\eqref{bcflow}
to identify flows which are visualized as `brane processes'.
For some models we collected the complete results in appendix~B.

\subsection{Minimal Models, A series}
The unitary minimal models can be constructed as diagonal cosets of
the form $\widehat{su} (2)_{k}\oplus \widehat{su} (2)_{1}/\widehat{su}
(2)_{k+1}$ with an integer $k\geq 1$. The modular invariant partition
functions for these models are completely
classified~\cite{Cappelli:1987hf,Cappelli:1987xt,Kato:1987td} (see
also \cite{FrancescoCFT}); in this
subsection we shall deal with the A series which is sometimes
denoted as $(A_{k+1},A_{k+2})$.

The sectors of the theory are labeled by three integers
$[l_{1},l_{2},l']$ in the range $l_{1}=0,\dots ,k; l_{2}=0,1; l'=0,\dots
,k+1$. Selection rules force the sum $l_{1}+l_{2}+l'$ to be even, and
there is an identification $[l_{1},l_{2},l']\sim
[k-l_{1},1-l_{2},k+1-l']$ between admissible labels\footnote{The
relation to the usual Kac labels $(r,s)$ is $r=l_{1}+1$ and $s=l'+1$.}. 
The adjoint field from the sector $[0,0;2]$ that induces the flow
described by the rule~\eqref{bcflow} has conformal weight $h=
(k+1)/ (k+3)$. 

In the A-series we are in the Cardy case, i.e.\ the boundary
conditions $\alpha $ are labeled by 
triples $[L_{1},L_{2},L ']$
taking values in the same range as the sectors 
including selection and identification rules.

The annulus coefficients are just given by the fusion rules $N^{(k)}$
and $N^{(k+1)}$ of
$\widehat{su} (2)_{k}$ and $\widehat{su} (2)_{k+1}$ resp.,
\[
n_{[l_{1},l_{2},l'][L _{1},L _{2},L ']}{}^{[J
_{1},J _{2},J ']}\ =\ N^{(k)}_{l_{1}L _{1}}{}^{J _{1}}\
N^{(k+1)}_{l'L '}{}^{J '}+  N^{(k)}_{k-l_{1}\, L _{1}}{}^{J _{1}}\
N^{(k+1)}_{k+1-l'\, L '}{}^{J '}\ \ .
\]

We now want to give a pictorial
representation of the boundary conditions. Coset models can also be
formulated as non-linear $\sigma $-models on a background geometry
which is essentially given by the space $G/\Ad \ H$ where we divide
the group $G$ by the adjoint action of the subgroup
$H$. The boundary conditions can then be described by certain
subspaces ('branes') onto which the boundary of our two-dimensional
world-sheet is mapped~\cite{Gawedzki:2001ye,Elitzur:2001qd}. One
should be aware that this geometrical interpretation is only valid for
large values of the level. If one, however, views the pictures just as a
nice tool to illustrate the boundary conditions, we can profitably 
employ them for arbitrary levels.

\begin{figure}
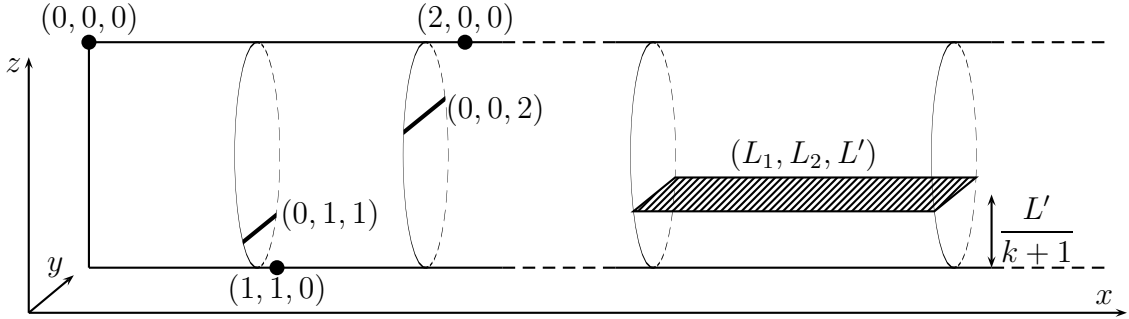

\begin{center}
\minimalmodelgeometryfinitek 
\end{center}
\caption{\label{minimalmodelgeometryfinitek} The geometrical
representation of boundary conditions in the minimal model A-series. }
\end{figure}

In the case of minimal models 
we describe the background as a solid cylinder with
squeezed ends~\cite{Fredenhagen:2001kw}. The boundary conditions are
represented by branes, extended objects of dimension 0,1 and 2. Let
$x$ be the coordinate along the axis of the cylinder, $z$ the
coordinate along the squeezed ends and $y$ a third coordinate
perpendicular to the others (see fig.\
\ref{minimalmodelgeometryfinitek}). All branes $[L_{1},L_{2},L']$ 
are located along surfaces of constant $z$ and are maximally extended in the
$y$-direction. In $x$ they stretch between two values $x_{\rm min}$ and
$x_{\rm max}$. The boundary conditions with $L'=0,k+1$ are
represented as points at the top or bottom of the cylinder, the ones
with $L=0,k$ are one-dimensional objects stretching in $y$-direction
(see fig.\ \ref{minimalmodelgeometryfinitek}).
We shall give explicit formulas for $z, x_{\rm min},
x_{\rm max}$. The coordinate $z$ lies in the range $[0,1]$, and the
coordinate $x$ takes values in $[0,k]$,
\[
z\ =\ \left\{\begin{array}{cl}
\frac{L'}{k+1} & \text{ for } L_{2}=1\\
1-\frac{L'}{k+1} & \text{ for } L_{2}=0
\end{array} \right. \ , \quad \begin{array}{rl}
x_{\rm min}\ =&  |L_{1}-\frac{k}{k+1}L'|\\
x_{\rm max}\ =&  \max \{ L_{1}+\frac{k}{k+1}L', 2k-L_{1}-\frac{k}{k+1}L' \}
\end{array}
\]

The rule~\eqref{bcflow} describes a large number of flows for many
different starting configurations. We shall concentrate here on two
types of starting points: a single boundary condition and a
superposition of boundary conditions of the form $[0,L
_{2},L ']$.

\begin{figure}
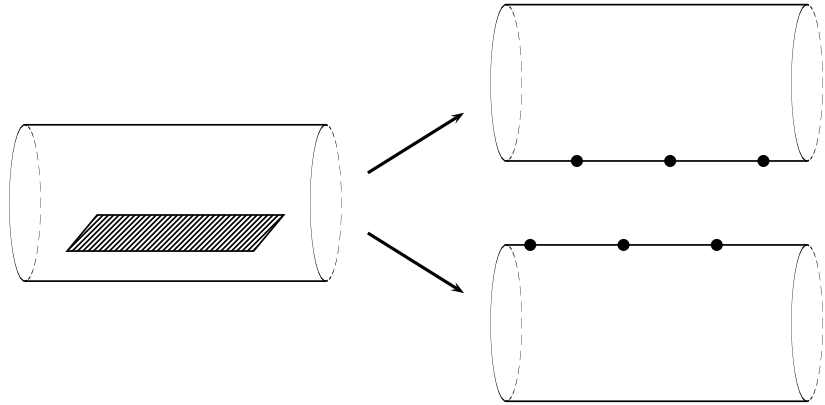

\begin{center}
\scalebox{.8}{\minmodsmallkprocesses}
\end{center}
\caption{\label{minmodsmallkprocessesI}A pictorial representation of
the flows~\eqref{mmAflowI} and~\eqref{mmAflowII} for
$L_{1}<L',L_{1}+L'\leq k$: a single brane can
flow to a superposition of point-like branes. }
\end{figure}

\begin{figure}
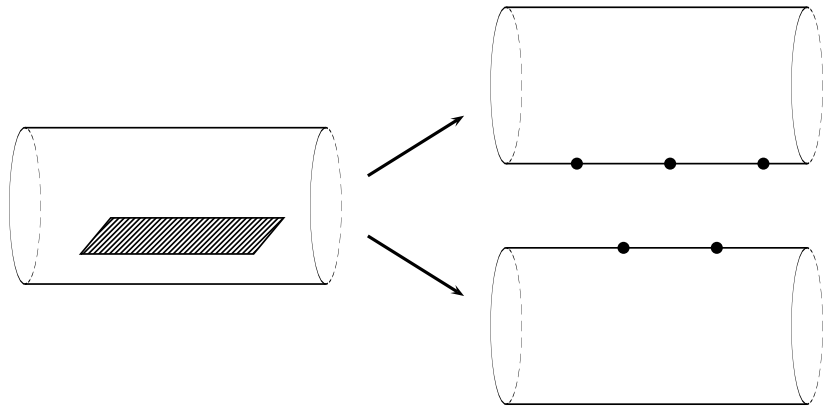

\begin{center}
\scalebox{.8}{\minmodsmallkprocessesII}
\end{center}
\caption{\label{minmodsmallkprocessesII}A pictorial representation of
the flows~\eqref{mmAflowI} and~\eqref{mmAflowII} for
$L_{1}\geq L',L_{1}+L'\leq k$: a single brane can
flow to a superposition of point-like branes.}
\end{figure}

Assume that we want to study flows starting from the boundary condition
$[L_{1},L_{2},L']$ with $1\leq L' \leq k$. To apply our
rule~\eqref{bcflow}, we have
to find a boundary condition $\alpha $ and a 'boundary spin' $S$ s.t.\ 
\[
(0,\creplarge{S}|_{\Lie{h}}) \hat{\times }\alpha \ =\ [L_{1},L_{2},L'] \ \ .
\] 
On the one hand we can set $\alpha = [L_{1},L_{2},0]$ and $S=
(L',0)$. This corresponds to the flow
\begin{equation}\label{mmAflowI}
[L_{1},L_{2},L']\ \longrightarrow \ \makebox[6cm][l]{$\displaystyle\bigoplus_{J}\
N^{(k)}_{L_{1}L'}{}^{J}\ [J,L_{2},0] \ \ .$}
\end{equation}
On the other hand, the choice $\alpha = [k-L_{1},1-L_{2},0]$ and $S=
(k+1-L',0)$ leads to the flow
\begin{equation}\label{mmAflowII}
[L_{1},L_{2},L']\ \longrightarrow \ \makebox[6cm][l]{$\displaystyle\bigoplus_{J}\
N^{(k)}_{L_{1}\, L'-1}{}^{J}\ [J,1-L_{2},0] \ \ .$}
\end{equation}
The number of elementary boundary conditions appearing on the r.h.s.\
of \eqref{mmAflowI} and \eqref{mmAflowII} can be different depending
on the values of $L_{1},L'$. Assume that $L_{1}+L'\leq k$ which we can
always achieve by using the identification rules. For $L_{1}<L'$ we
find a superposition of $L_{1}+1$ boundary conditions 
in both flows \eqref{mmAflowI} and \eqref{mmAflowII}. These flows are
illustrated in fig.\ \ref{minmodsmallkprocessesI}. For $L_{1}\geq
L'$ there are $L'+1$ boundary conditions on the r.h.s.\ of
\eqref{mmAflowI}, but in \eqref{mmAflowII} we find a superposition of 
$L'$ boundary conditions (see fig.\ \ref{minmodsmallkprocessesII}).

The first of the flows, \eqref{mmAflowI}, 
has been analyzed in perturbation theory for
large $k$ in~\cite{Recknagel:2000ri}, the second one,
\eqref{mmAflowII}, cannot be seen in
this limit. Nevertheless, both flows are known to
exist~\cite{Chim:1996kf,Ahn:1998xm,Lesage:1998qf}. They are generated
by the $[0,0,2]$ field (in Kac labels $(1,3)$) and differ by the sign
of the perturbation. This is in agreement with our general
statements~\eqref{list} on the boundary fields generating the flow.
\smallskip

\begin{figure}
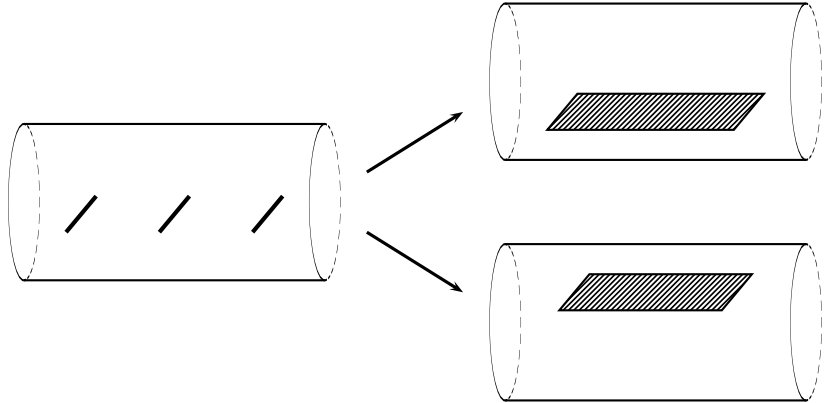

\begin{center}
\scalebox{.8}{\minmodsmallkprocessesIII}
\end{center}
\caption{\label{minmodsmallkprocessesIII}Pictorial representation of
the flows~\eqref{mmAflowIII} and~\eqref{mmAflowIV}: a superposition of
string-like branes can flow to a single brane.}
\end{figure}

Now let us choose a superposition of boundary conditions with a
trivial first label. We set $S= (L_{1},0)$ $(1\leq L_{1}\leq k)$ and $\alpha =
[0,L_{2},L']$ in~\eqref{bcflow} and obtain
\begin{equation}\label{mmAflowIII}
\bigoplus_{} \ N^{(k+1)}_{L_{1}L' }{}^{J} \ [0,L_{2},J]
\longrightarrow \ \makebox[5cm][l]{$\displaystyle [L_{1},L_{2},L'] \ \
.$}
\end{equation}
On the other hand, we could choose $\alpha = [0,L_{2},k+1-L']$ and
$S= (k+1-L_{1},0)$ which leads to
\begin{equation}\label{mmAflowIV}
\bigoplus_{} \ N^{(k+1)}_{L_{1}L' }{}^{J} \ [0,L_{2},J]
\longrightarrow \ \makebox[5cm][l]{$\displaystyle [L_{1}-1,1-L_{2},L']
\ \ .$}
\end{equation}
The two flows are illustrated in fig.\ \ref{minmodsmallkprocessesIII}. 
Again, perturbation theory for large $k$ can only see the first of
these flows~\cite{Fredenhagen:2001kw,Graham:2001pp}.

\subsection{Critical Ising model}
\begin{figure}
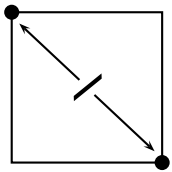

\begin{center}
\isingmodelflows 
\end{center}
\caption{\label{isingmodelflows}Ising model: flows from the free
boundary condition to spin up or spin down.}
\end{figure}

The simplest model in the unitary minimal A-series is the critical Ising
model with $k=1$. There are three boundary conditions, the free one
and two with fixed spin (up or down) at the boundary. Our geometrical
picture reduces to a cushion-like background where the fixed boundary
conditions are point-like objects at the top and bottom whereas the
free boundary condition is a string-like object sitting precisely in
the middle of the cushion (see fig.\ \ref{isingmodelflows}). 

Starting from the free condition, the system can be driven into a
theory with fixed spin. These are precisely the two
flows~\eqref{mmAflowI}, \eqref{mmAflowII}. They are depicted in 
fig.~\ref{isingmodelflows}. Flows starting from a superposition of boundary
conditions can be found in appendix~B.

\subsection{Tricritical Ising model}
\begin{figure}
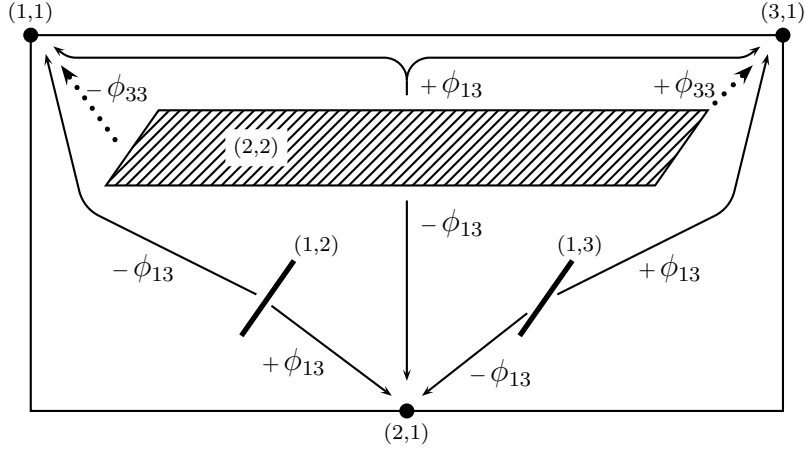

\begin{center}
\tricriticalisingmodel
\end{center}
\caption{\label{tricriticalisingmodel}Boundary RG flows in the
tricritical Ising model induced by the $(1,3)$-field ($\phi _{13}$)
and the $(3,3)$-field ($\phi _{33}$).}
\end{figure}

The second model in the unitary minimal series is the tricritical
Ising model with central charge $c=7/10$. Once more, the
flows triggered by the $(1,3)$-field as analyzed in~\cite{Chim:1996kf} 
are correctly reproduced by~\eqref{mmAflowI} and
\eqref{mmAflowII}. There are, however, more flows known which
correspond to a perturbation with other
fields~\cite{Affleck:2000jv}. As the rule depends on the specific
coset construction, it is possible to find additional flows by
choosing different coset realizations of the same theory. 
For the tricritical 
Ising model, such alternative realizations do exist. One is 
given by $(E_{7})_{1}\oplus(E_{7})_{1}/(E_{7})_{2} $. 
When we apply our rule to this coset construction, it reproduces the two
known flows caused by the $(3,3)$-field. In Kac labels they read 
\[
(2,2)\ \longrightarrow\ (3,1)\ ,\ \ (2,2)\ \longrightarrow\ (1,1)\ ,
\]
and they are depicted together with the other flows in
fig.\ \ref{tricriticalisingmodel}.  
These two flows also appear in higher minimal models \cite{Graham:2001tg} 
where we do not know a coset realization for the
$(3,3)$-perturbations. 
This may be related to the observation that the tricritical Ising model 
seems to be the only theory in which the considered perturbations are 
integrable \cite{Graham:2001tg}. Nevertheless, recovering flows from
the exceptional $E_{7}$ coset construction can be considered as an
important check of the conjectured rule.

There are more realizations of the tricritical Ising model as coset
model~\cite{Bowcock:1987mw}, 
but only for one of them our rule predicts flows starting from single
boundary conditions. This is the construction 
as a $so (7)_{1}/(G_{2})_{1}$ coset model. The flows found there
coincide with the $(1,3)$-flows~\eqref{mmAflowII}, i.e.\  with
those flows found in the $SU (2)$ construction that cannot be seen
in the perturbative approach. 

In appendix~B we collected the complete results, i.e.\ all flows
described by the rule~\eqref{bcflow} including those that start from a
superposition of boundary conditions, for the $su (2)$ and
the $E_{7}$ construction. 

\subsection{Minimal Models, D-Series}
For the minimal models with $k\geq 3$, there is in addition to the
diagonal modular invariant (A-type) another modular invariant giving
rise to the D-series of minimal models. Up to some exceptional values
of $k$, these form all possible modular invariants for the minimal models.
Depending on $k$ being even or odd, we distinguish between minimal
models of type $({\rm D}_{\frac{k+4}{2}},{\rm A}_{k+2})$ and
$({\rm A}_{k+1},{\rm D}_{\frac{k+5}{2}})$, 
and we shall discuss these two classes
of models separately.

\subsubsection*{(D,A): $\boldsymbol{k}$ even}
Boundary conditions $\alpha $ in the (D,A) models are labeled by triples 
$[L_{1},L_{2},L']$ where $L'=0,\dots ,k+1$ and $L_{2}=0,1$ lie in the
usual ranges whereas $L_{1}$ takes the values $0,1,\dots
,\frac{k}{2}-1,[\frac{k}{2},+],[\frac{k}{2},-]$. The sum
$L_{1}+L_{2}+L'$ of 
boundary labels has to be even\footnote{When we use $L_{1}$ only as a
numerical value and not as a label, 
we forget about the possible signs $\pm $ from
fixed-point resolution.}. We have the identifications
$[L_{1},L_{2},L']\sim [\hat{L}_{1},1-L_{2},k+1-L']$ where
\begin{equation}\label{defofhat}
\hat{L}_{1}\ = \left\{\begin{array}{cl}
[\frac{k}{2},\mp] & \text{ for } L_{1}=[\frac{k}{2},\pm] \text{ and
$\frac{k}{2}$ odd}\\
L_{1} & \text{ otherwise .}
\end{array} \right.
\end{equation}
The annulus coefficients can be written as a combination of the
annulus coefficients $n^{{\rm D},k}$ for the $\widehat{su} (2)$ model at level $k$ with
D-type partition function and the fusion rules of $\widehat{su}
(2)_{1}$ and $\widehat{su}(2)_{k+1}$,
\[
n_{[l_{1},l_{2},l'] [L_{1},L_{2},L']}^{\quad \quad \quad \quad  [J_{1},J_{2},J']}\ =\ 
n^{{\rm D} ,k}_{l_{1}L_{1}}{}^{J_{1}}\, N^{(1)}_{l_{2}L_{2}}{}^{J_{2}}\,
N^{(k+1)}_{l'L'}{}^{J'} \ +\ n^{{\rm D},k}_{l_{1}\hat{L}_{1}}{}^{J_{1}}\, 
N^{(1)}_{l_{2} (1-L_{2})}{}^{J_{2}}\, N^{(k+1)}_{l' (k+1-L')}{}^{J'}\
\ .
\]
Here, $n^{{\rm D},k}$ is given by
\begin{equation}\label{anncoeffD}
n^{{\rm D},k}_{l_{1}L_{1}}{}^{J_{1}}\ =\ \left\{\begin{array}{ll}
N^{(k)}_{l_{1}L_{1}}{}^{J_{1}}+N^{(k)}_{(k-l_{1})L_{1}}{}^{J_{1}} &
\text{ for } J_{1},L_{1}\not= \frac{k}{2}\parbox[c][8mm][c]{0cm}{}\\
N^{(k)}_{l_{1}\frac{k}{2}}{}^{J_{1}}&
\text{ for } J_{1}\not= \frac{k}{2},\ L_{1}=[\frac{k}{2},\pm ]\parbox[c][8mm][c]{0cm}{}\\
N^{(k)}_{l_{1}L_{1}}{}^{\frac{k}{2}} &
\text{ for } J_{1}=[\frac{k}{2},\pm ],\ L_{1}\not= \frac{k}{2}\parbox[c][8mm][c]{0cm}{}\\
\delta _{l_{1}\!\!\!\mod 4}&
\text{ for } J_{1}=L_{1} = [\frac{k}{2},\pm ]\parbox[c][8mm][c]{0cm}{}\\
\delta _{l_{1}-2\!\!\! \mod 4}&
\text{ for } J_{1}= [\frac{k}{2},\pm],\ L_{1}=[\frac{k}{2},\mp]\parbox[c][8mm][c]{0cm}{}
\end{array} \right.
\end{equation}
\smallskip

\begin{figure}
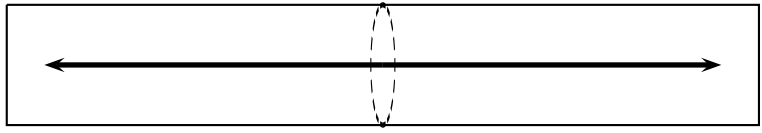

\begin{center}
{\minimalmodelDA} 
\end{center}
\caption{\label{minimalmodelDA}The geometry of the (D,A)-minimal model
is the cylinder of the A-minimal model modulo the reflection
(indicated by the arrow) at the
$x=k/2$-plane.}
\end{figure}

The geometry of the minimal models of the (D,A)-series is the cylinder
of the A-series divided by the reflection at the plane at $x=k/2$ (see
fig.\ \ref{minimalmodelDA}).
The branes which are symmetric with respect to the reflection split
into two `fractional branes'.
\smallskip

We are only going to discuss flows starting from a single boundary
condition $[L_{1},L_{2},L']$ with $L'\not= 0,k+1$. Because of the
identification rules we are allowed to choose $L'\leq \frac{k}{2}$.
We distinguish three cases:
\begin{itemize}
\item $L_{1}+L'<\frac{k}{2}$:

In~\eqref{bcflow} we choose $\alpha = [L_{1},L_{2},0]$ and $S= (L',0)$
and find the flow
\begin{equation}\label{mmDAflowI}
[L_{1},L_{2},L']\ \longrightarrow \ \makebox[6cm][l]{$\displaystyle\bigoplus_{J}
N^{(k)}_{L_{1}L'}{}^{J}\, [J,L_{2},0] \ \ .$}
\end{equation}
Alternatively, we choose $\alpha = [L_{1},1-L_{2},0]$ and $S=
(k+1-L',0)$ and obtain
\begin{equation}\label{mmDAflowII}
[L_{1},L_{2},L']\ \longrightarrow \ \makebox[6cm][l]{$\displaystyle \bigoplus_{J}
N^{(k)}_{L_{1},(L'-1)}{}^{J}\, [J,1-L_{2},0] \ \ .$}
\end{equation}
\begin{figure}
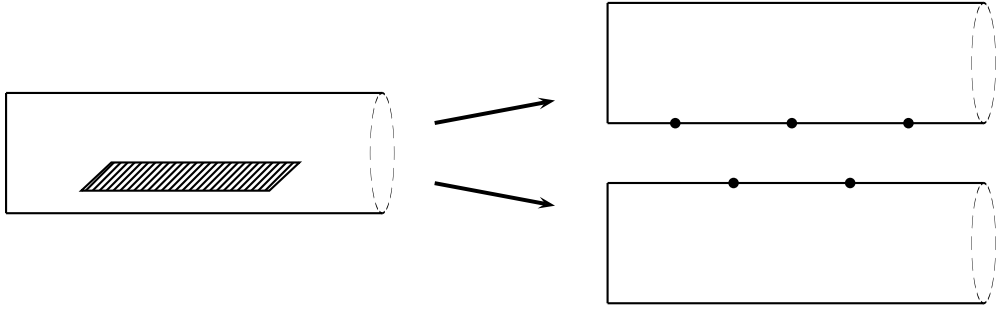

\begin{center}
\minmodflowsDAI 
\end{center}
\caption{\label{minmodflowsDAI}The flows~\eqref{mmDAflowI}
and~\eqref{mmDAflowII} illustrated in the half-cylinder geometry of
the (D,A)-minimal models.}
\end{figure}
The flows are illustrated in fig.\ \ref{minmodflowsDAI}.

\item $\frac{k}{2}\leq L_{1}+L',\ L_{1}\not= \frac{k}{2}$

As in the previous case, we set $\alpha = [L_{1},L_{2},0]$ and $S=
(L',0)$. The rule~\eqref{bcflow} then leads to
\begin{equation}\label{mmDAflowIII}
[L_{1},L_{2},L'] \ \longrightarrow \ \makebox[8cm][l]{$\displaystyle\begin{array}[t]{l}\displaystyle \bigoplus_{J< \frac{k}{2}}
\bigl( N^{(k)}_{L_{1}L'}{}^{J}+N^{(k)}_{L_{1}L'}{}^{k-J} \bigr) [J,L_{2},0]\\
 \quad \oplus \quad N^{(k)}_{L_{1}L'}{}^{\frac{k}{2}} \ ([\frac{k}{2},+],L_{2},0) \\[2mm]
 \quad \oplus \quad N^{(k)}_{L_{1}L'}{}^{\frac{k}{2}} \ ([\frac{k}{2},-],L_{2},0)
\end{array}$ }
\end{equation}
We find a second flow for $\alpha = [L_{1},1-L_{2},0]$ and $S= (k+1-L')$,
\begin{equation}\label{mmDAflowIV}
[L_{1},L_{2},L'] \ \longrightarrow \ \makebox[8cm][l]{$\displaystyle\begin{array}[t]{l}\displaystyle \bigoplus_{J< \frac{k}{2}}
\bigl( N^{(k)}_{L_{1} (L'-1)}{}^{J}+N^{(k)}_{L_{1} (L'-1)}{}^{k-J} \bigr) [J,1-L_{2},0]\\
 \quad \oplus \quad N^{(k)}_{L_{1} (L'-1)}{}^{\frac{k}{2}} \ [[\frac{k}{2},+],1-L_{2},0] \\[2mm]
 \quad \oplus \quad N^{(k)}_{L_{1} (L'-1)}{}^{\frac{k}{2}} \ ([[\frac{k}{2},-],1-L_{2},0]
\end{array}$ }
\end{equation}
\begin{figure}
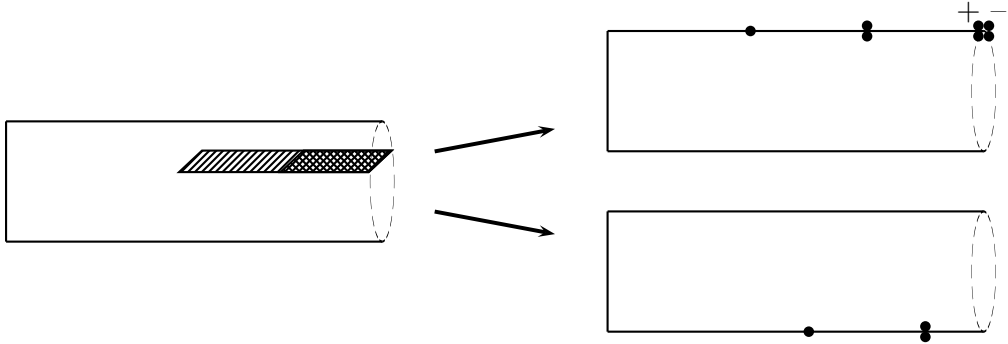

\begin{center}
\minmodflowsDAII 
\end{center}
\caption{\label{minmodflowsDAII}The flows~\eqref{mmDAflowIII}
and~\eqref{mmDAflowIV} in the (D,A)-minimal model. Some of the branes
appear with multiplicity 2. Note that the
point-like branes sitting on the fixed-plane come in pairs of
fractional branes indicated by $+$ and $-$.}
\end{figure}
We have depicted the flows in fig.\ \ref{minmodflowsDAII}.

\item $L_{1}=[\frac{k}{2},+]$ \hspace*{1cm} (analogously for
$[\frac{k}{2},-]$)

Here we find the flows
\begin{equation}\label{mmDAflowV}
\hspace*{-2cm}[L_{1},L_{2},L']\  \longrightarrow \ \makebox[10cm][l]{$
\displaystyle \bigoplus_{J< \frac{k}{2}}\ 
N^{(k)}_{L'\,\frac{k}{2}}{}^{J}\ [J,L_{2},0]\ \oplus \
\left\{ 
\begin{array}{cl}
([ \frac{k}{2} , + ],L_{2},0) & L'=0 \mod 4 \\[1mm]
{( [ \frac{k}{2} , - ],L_{2},0)} & L'=2 \mod 4 \\[1mm] 
0 & L' \mbox{ odd}
\end{array} 
\right. 
$ }
\end{equation}
and
\begin{equation}\label{mmDAflowVI}
\hspace*{-2cm}[L_{1},L_{2},L']\  \longrightarrow \ \makebox[10cm][l]{$
\displaystyle \bigoplus_{J< \frac{k}{2}}\ 
N^{(k)}_{(L'-1)\,\frac{k}{2}}{}^{J}\ [J,L_{2},0]\ \oplus \
\left\{ 
\begin{array}{cl}
([ \frac{k}{2} , + ],L_{2},0) & L'=1 \mod 4 \\[1mm]
{ ([ \frac{k}{2} , - ],L_{2},0)} & L'=3 \mod 4 \\[1mm] 
0 & L' \mbox{ even}
\end{array} 
\right. 
$ }
\end{equation}
One example for the described flows with $L'=2$ can be found in
figure~\ref{minmodflowsDAIII}. 
\begin{figure}
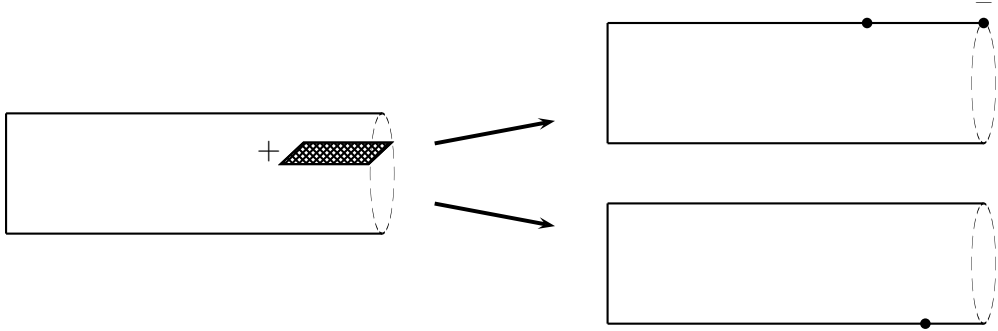

\begin{center}
\minmodflowsDAIII 
\end{center}
\caption{\label{minmodflowsDAIII} The flows~\eqref{mmDAflowV}
and~\eqref{mmDAflowVI} for $L'=2$. In contrast to the flows in
figure~\ref{minmodflowsDAII} all branes only appear with multiplicity
1. Note that the sign of the fractional brane at the RG end-point
depends on the value of $L'$.}
\end{figure}
\end{itemize}

\subsubsection*{(A,D): $\boldsymbol{k}$ odd}
In the (A,D)-series of the minimal model, we label the boundary
conditions by triples $[L_{1},L_{2},L']$ where $L_{1}=0,\dots ,k$, 
$L_{2}=0,1$, and $L'=0,\dots
,\frac{k-1}{2},[\frac{k+1}{2},+],[\frac{k+1}{2},-]$. Selection rules
force the sum $L_{1}+L_{2}+L'$ to be even, triples $[L_{1},L_{2},L']$
and $[k-L_{1},1-L_{2},\hat{L}']$ are identified. Here, $\hat{L}'$ is
defined as in~\eqref{defofhat} with $k$ replaced by $k+1$.

The annulus coefficients are given by the fusion rules of
$\widehat{su} (2)$ at level $k$ and $1$ and by the annulus
coefficients $n^{{\rm D},k+1}$ of the $\widehat{su} (2)_{k+1}$ model
with D-type modular invariant (see \eqref{anncoeffD}),
\[
n_{[l_{1},l_{2},l'] [L_{1},L_{2},L']}^{\quad \quad \quad \quad [J_{1},J_{2},J']}\ =\ 
N^{(k)}_{l_{1}L_{1}}{}^{J_{1}}\, N^{(1)}_{l_{2}L_{2}}{}^{J_{2}}\,
n^{{\rm D},k+1\ }_{l'L'}{}^{J'} \ +\ N^{(k)}_{l_{1} (k-L_{1})}{}^{\!J_{1}}\, 
N^{(1)}_{l_{2} (1-L_{2})}{}^{\!J_{2}}\, n^{{\rm D},k+1\ }_{l'
\hat{L}'}{}^{J'} \ \ .
\] 

\begin{figure}
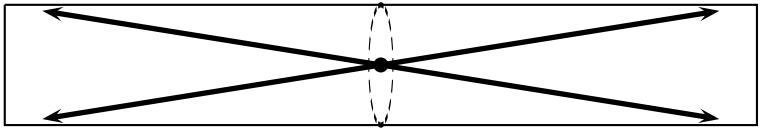

\begin{center}
{\minimalmodelAD} 
\end{center}
\caption{\label{minimalmodelAD}The (A,D)-minimal model is described by
the cylinder geometry of the A-series modulo the reflection at the
center (indicated by the arrows).}
\end{figure}

The geometry of the (A,D)-minimal models is obtained from the cylinder
geometry of the A-series by dividing out the reflection at the center
(see fig.\ \ref{minimalmodelAD}). 
A brane which is symmetric under this reflection
splits into two fractional branes.
\smallskip

As in the (D,A)-case we look for flows starting from a single boundary
condition $[L_{1},L_{2},L']$ with $L'\not= 0$. 
We have to distinguish two cases:
\begin{itemize}
\item $L'\not= \frac{k+1}{2}$

When we choose $\alpha = [L_{1},L_{2},0]$ and $S= (L',0)$
in~\eqref{bcflow}, we find
\begin{equation}\label{mmADflowI}
[L_{1},L_{2},L']\ \longrightarrow\ \makebox[6cm][l]{$
\displaystyle \bigoplus_{J}\ N^{k}_{L_{1}L'}{}^{J}\ [J,L_{2},0] \ \ .$}
\end{equation}
Similarly, for $\alpha = [k-L_{1},1-L_{2},0]$ and $S= (k+1-L',0)$ we
obtain
\begin{equation}\label{mmADflowII}
[L_{1},L_{2},L']\ \longrightarrow\ \makebox[6cm][l]{$
\displaystyle \bigoplus_{J}\ N^{k}_{L_{1} (L'-1)}{}^{J}\ [J,1-L_{2},0] \ \ .$}
\end{equation}
\begin{figure}
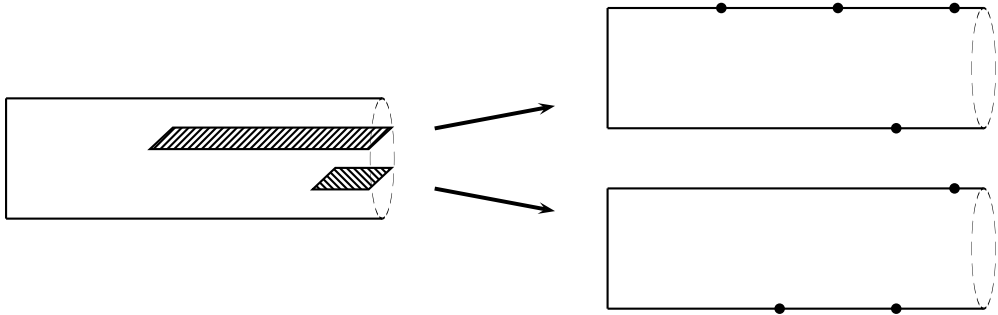

\begin{center}
{\minmodflowsADI}
\end{center}
\caption{\label{minmodflowsADI}The flows~\eqref{mmADflowI}
and~\eqref{mmADflowII} in the (A,D)-minimal model.}
\end{figure}
These flows are shown in fig.\ \ref{minmodflowsADI}. 

\item $L'=[\frac{k+1}{2},\pm ]$ 

By setting $\alpha = [L_{1},L_{2},0]$ and $S= (\frac{k+1}{2},0)$ we 
find a flow for a superposition of two boundary conditions,
\begin{equation}\label{mmADflowIII}
[L_{1},L_{2},[{\textstyle \frac{k+1}{2}},+]]\oplus
[L_{1},L_{2},[{\textstyle \frac{k+1}{2}},-]]\
\longrightarrow \ \bigoplus_{J}\ N^{(k)}_{L_{1}\frac{k+1}{2}}{}^{J}\ 
[J,L_{2},0]\ \ .
\end{equation}
An example for this flow is shown in fig.\ \ref{minmodflowsADII}.
\begin{figure}[!t]
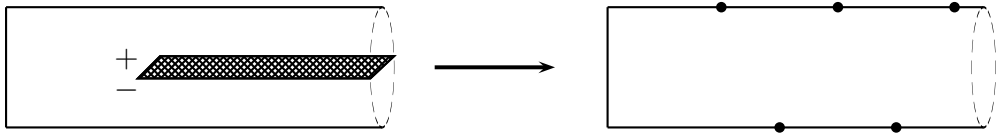

\begin{center}
{\minmodflowsADII} 
\end{center}
\caption{\label{minmodflowsADII}The flow~\eqref{mmADflowIII} in the
(A,D)-minimal models starting from a superposition of two fractional
branes $+$ and $-$.}
\end{figure}
\end{itemize}

\subsection{Parafermion series}
The parafermion series can be realized by the cosets $\widehat{su}
(2)_{k}/\widehat{u}_{2k}$. The sectors of the theory are labeled by
pairs $[l,l']$ where $l=0,\dots ,k$ and $l'$ is a $2k$-periodic
integer for which we usually choose the range 
$l'=-k+1,\dots ,k$. Selection
rules force the sum $l+l'$ to be even, and the pairs $[l,l']$ and
$[k-l,l'+k]$ are identified. The fields which appear as
perturbing fields in our rule have conformal weight $h= (k-1)/k$.

The maximally symmetric branes in parafermion theories come in two
classes: the untwisted branes (A-branes), and the twisted branes
(B-branes). The untwisted branes are the usual Cardy branes and carry
labels $[L,L']$ from the same set as the sectors. The annulus
coefficients are given by
\[
n_{[l,l'][L,L']}{}^{[J,J']}\ =\ \delta _{l'+L'-J'\!\!\!\mod 2k}
\ N^{(k)}_{lL}{}^{J}\ +\ \delta _{l'+L'-J'+k\!\!\!\mod 2k}
\ N^{(k)}_{l (k-L)}{}^{J}\ \ 
\]
where $N^{(k)}$ are the fusion rules of $\widehat{su} (2)_{k}$.
\smallskip

\begin{figure}
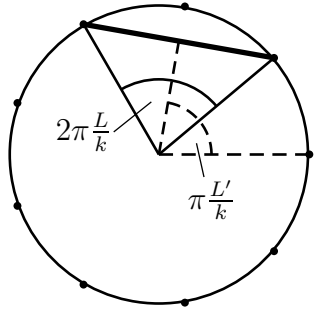

\begin{center}
\parafermionbranemod 
\end{center}
\caption{\label{parafermionbrane} A generic untwisted brane $[L,L']$
in the parafermion model and the geometric interpretation of the
labels of the brane. The possible positions of the point-like branes
of type $[0,L']$ are also indicated.}
\end{figure}

In the limit of large levels $k$, the parafermion models can be
described by a non-linear $\sigma $-model on a disc with non-trivial
metric (sometimes this geometry is called 'bell'). We want to use this
picture to visualize boundary conditions as branes. 
The untwisted branes $[0,L']$ appear as
point-like objects sitting at $k$ special equidistant points on the
boundary of the circle. The other untwisted branes $[L,L']$ are
one-dimensional objects that stretch between these points (see fig.\
\ref{parafermionbrane}). These pictures have been introduced
in~\cite{Maldacena:2001ky}.
\smallskip

\begin{figure}
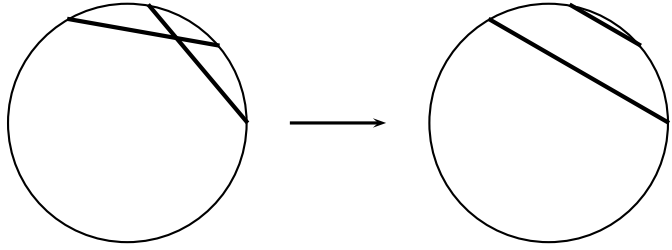

\begin{center}
\scalebox{.8}{\paraprocI} 
\end{center}
\caption{\label{paraproc1}Two untwisted branes in the parafermion
model with $L=2$ flow to a configuration of a $L=1$ and a $L=3$ brane.}
\end{figure}

\begin{figure}[!t]
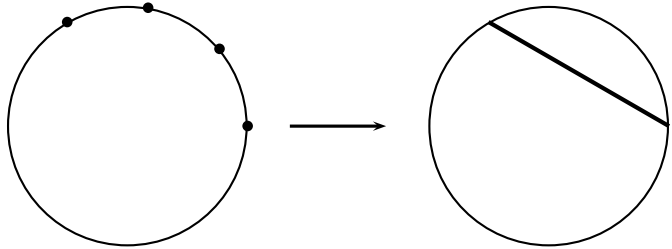

\begin{center}
\scalebox{.8}{\parafermionprocess} 
\end{center}
\caption{\label{paraprocII}Four point-like branes condense into a
single extended brane.}
\end{figure}

Let us now apply the rule~\eqref{bcflow} to untwisted boundary
conditions. The general result is 
\[
[L,L'-S]\oplus [L,L'-S+2]\oplus \cdots \oplus [L,L'+S]\
\longrightarrow\ \bigoplus_{J}\, N^{(k)}_{SL}{}^{J}[J,L']
\]
for any label $L,L',S$ with $L+L'+S$ even. An example of a flow with
$L=2,S=1$ is graphically presented in fig.\ \ref{paraproc1}.
Particularly interesting is the case $L=0$ where the end configuration
consists only of a single boundary condition:
\[
[0,L'-S]\oplus [0,L'-S+2]\oplus \cdots \oplus [0,L'+S]\
\longrightarrow\  [S,L'] \ \ .
\]
Such a flow is shown in fig.\ \ref{paraprocII}.
\medskip

In addition to the untwisted (Cardy) boundary conditions 
there are twisted ones which involve a non-trivial automorphism
$\omega $. In the $u (1)$ part it acts as reflection,
\[
\omega \bigl( J \bigr) (z)\ =\ -J (z) \ \ ,
\]
on the numerator $su (2)$ it acts only as an inner automorphism. The
twisted boundary conditions have first been constructed
in~\cite{Maldacena:2001ky}.
\begin{figure}
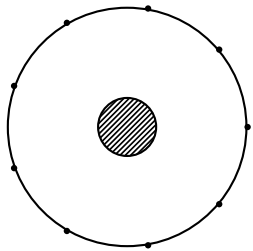

\begin{center}
\scalebox{.8}{\parafermiontwbrane} 
\end{center}
\caption{\label{parafermiontwbrane} A generic twisted brane $[L,\pm ]$
in the parafermion model. For $L=0$ the brane becomes point-like. If
$k$ is even we find two branes for
$L=\frac{k}{2}$ which cover the whole disc.}
\end{figure} 

These boundary conditions are labeled by pairs $[L,L']$ where $L=0,\dots ,k$
is an integer coming from the numerator part, and the sign $L'=\pm $
comes from the twisted $U(1)$. Selection rules force $L$ to be even 
in combination with the
sign $L'=+$, and odd if it comes with $L'=-$. As $L'$ is 
determined by $L$, we shall often leave it out and write $[L,\cdot ]$.
Furthermore, there is an identification
between pairs, $[L,+]\sim [k-L,(-1)^{k}]$. For even $k$, the pair
$[\frac{k}{2},L' ]$ is a fixed-point of this identification,
and the corresponding boundary condition has to be resolved into two
elementary boundary conditions $[\frac{k}{2},L' ;\pm ]$.

The annulus coefficients (before fixed-point resolution) are given by 
\[
n_{[l,l'][L,L']}{}^{[J,J']}\ =\ N^{(k)}_{lL}{}^{J}n_{l'L'}{}^{J'}+
N^{(k)}_{k-l\, L}{}^{J}n_{k+l'\,L'}{}^{J'}
\]
where the coefficients ${n_{S'L '}}^{J'}$ for the twisted $U
(1)$ read
\[
{n_{l'-}}^{-}\ =\ {n_{l'+}}^{+}\ =\ \left\{ 
\begin{array}{cl}1&l'\text{ even}\\
0&l'\text{ odd} \end{array}
\right.\ , \quad 
{n_{l'+}}^{-}\ =\ {n_{l'-}}^{+}\ =\ \left\{ 
\begin{array}{cl}0&l'\text{ even}\\
1&l\text{ odd} \end{array}
\right. \ .
\]
The resolution of the fixed-point for $L=k/2$ is straightforward.

In our geometric picture
the brane $[0,+]$ appears as point-like object in the center of the
disc, and the branes $[L,\cdot ]$ are 
two-dimensional discs placed at the
origin (see fig.\ \ref{parafermiontwbrane})\footnote{Note that from 
the point of view of
closed strings as derived in \cite{Maldacena:2001ky}, 
the smallest  twisted brane is not point-like, but has a small, non-zero
radius. We do not want to discuss these differences any further, as
the only purpose of the geometrical pictures here is to visualize the
boundary conditions and the boundary RG flows.}.

\begin{figure}
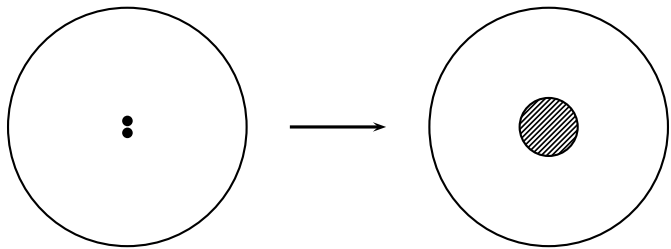

\begin{center}
\scalebox{.8}{\paratwproc}
\end{center}
\caption{\label{paratwproc}Two of the smallest twisted branes condense
into a larger disc.}
\end{figure}

The rule~\eqref{bcflow} applied to the twisted parafermion branes 
(ignoring again the fixed-point resolution) describes a flow from a
superposition of $S+1$ identical boundary conditions $[L,\cdot ]$ to
some other configuration, 
\[
(S+1)\ [L,\cdot ]\ \longrightarrow \bigoplus_{J}
N^{(k)}_{SL}{}^{J}[J,\cdot ] \ \  .
\]
An example for $L=0,S=1$ is shown in fig.\ \ref{paratwproc}.

There is a different realization of the parafermion series, namely 
as diagonal coset models $\widehat{su} (k)_{1}\oplus \widehat{su}
(k)_{1}/\widehat{su} (k)_{2}$. Here, the adjoint field which induces
the flows has conformal
weight $h=2/ (k+2)$. In this realization, we can even find flows
starting from single boundary conditions, e.g.\ a one-dimensional
brane at the boundary of the disc flows to a point-like one. We leave
it at these general words here, but we shall
work out the flows for $k=3$ in section~4.6.

\subsection{3-state Potts model}
The 3-state Potts model is a square lattice model where at each site
$i$ there is an angular variable $\theta_{i}$ 
taking values $0,\pm 2\pi /3$. The interaction is given by the
classical Hamiltonian
\[
\beta H\ = \ -c \sum_{\langle i,j\rangle }\cos
(\theta_{i}-\theta_{j})\ \ ,
\]
the sum running over nearest neighbor pairs. When the model is at its
critical coupling, it can be described by a conformal field
theory. Introducing a boundary into the problem, one can show that
there are 8 possible boundary
conditions~\cite{Affleck:1998nq,Fuchs:1998qn}. These are the free
boundary condition, the three different fixed boundary conditions,
three mixed boundary conditions (one of the three spin states is forbidden
at the boundary) and one additional boundary condition whose
interpretation in the classical Potts model is not as simple as for
the others (see~\cite{Affleck:1998nq} for details). We use the
nomenclature of~\cite{Affleck:1998nq} and call the boundary conditions
$F$, $A$, $B$, $C$, $AB$, $BC$, $AC$ and $N$ (for `new'), respectively.

The CFT describing the critical 3-state Potts model is a minimal model
of central charge $c=4/5$. It can be obtained by various coset
constructions: it belongs e.g.\ to the minimal D-series for $k=3$ and
to the parafermion series also for $k=3$. 
In addition to these two realizations, we shall review the construction
as a diagonal $su (3)$ coset. In all these realizations we determine flows
between boundary conditions using the rule~\eqref{bcflow}. In this
section, we shall see the rule in action in examples with twisted 
boundary conditions.
\begin{table}
\renewcommand{\arraystretch}{1.2}
\begin{center}
\begin{tabular}{ccccc}
\multicolumn{3}{c}{Boundary label from} & g-factor & Notation \\
$\frac{\widehat{su} (2)_{3}}{\widehat{u} (1)_{3}}$ & 
$\frac{\widehat{su} (2)_{3}\oplus \widehat{su}(2)_{1}}{\widehat{su}(2)_{4}}$ 
& $\frac{\widehat{su} (3)_{1}\oplus \widehat{su} (3)_{1}}{\widehat{su}
(3)_{2}}$ &  & from \cite{Affleck:1998nq}\\
\hline
$[0,0]$ &$[0,0,0]$ & $[(0,0),(0,0),(0,0)]$&$N$ & $A$\\
$[0,2]$  &$[0,0,2+]$ & $[(0,0),(0,1),(2,0)]$&$N$ & $B$\\
$[0,-2]$  &$[0,0,2-]$ & $[(0,0),(1,0),(0,2)]$&$N$ & $C$\\
$[1,1]$  &$[2,0,2-]$ & $[(0,0),(1,0),(1,0)]$&$N\lambda ^{2}$ & $AB$\\
$[1,3]$  &$[2,0,0]$ & $[(0,0),(0,0),(1,1)]$&$N\lambda ^{2}$ & $BC$\\
$[1,-1]$  &$[2,0,2+]$ & $[(0,0),(0,1),(0,1)]$&$N\lambda ^{2}$ & $AC$\\
$[1,-]$  &$[1,0,1]$ & $[0,0,0;\omega ]$ &$N\lambda ^{2}\sqrt{3}$ & $N$\\
$[0,+]$  &$[3,0,1]$ & $[0,0,1;\omega ]$ &$N\sqrt{3}$ & $F$ 
\end{tabular}
\end{center}
\caption{\label{potts}Boundary conditions in the 3-state Potts
model in three different coset constructions. 
The g-factors are given in
terms of $N^{4}= (5-\sqrt{5})/2$ and $\lambda ^{2}= (1+\sqrt{5})/2$.}
\end{table}
\smallskip

\begin{figure}
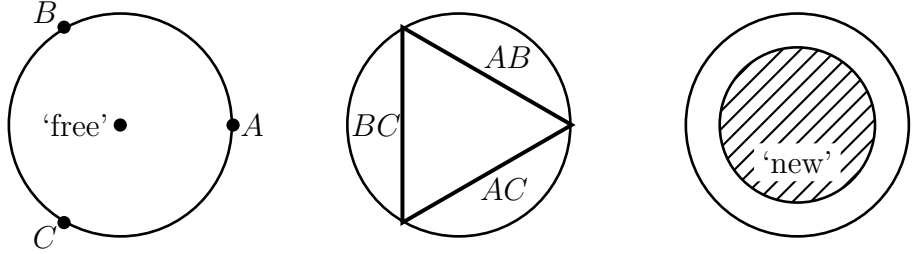

\begin{center}
{\pottsgeometry}
\end{center}
\caption{\label{pottsgeometry}Pictorial representation of boundary
conditions in the 3-state Potts model.}
\end{figure}

We start with the construction as a 
\[
\frac{\widehat{su} (2)_{3}}{\widehat{u}_{6}}
\]
coset that we already encountered in the discussion of parafermion
theories in section~4.5. The untwisted branes are labeled by pairs
$[L,L']$ where the labels $L$ and $L'$ lie in the range $L=0,1,2,3$ and
$L'=-2,-1,0,1,2,3$. Selection rules force the sum $L+L'$ to be even, and
the pairs $[L,L']$ and $[3-L,L'\pm 3]$ label the same brane. These are
the usual Cardy branes, and there are six of them in the model. 
We adopt the geometric interpretation from
section~4.5. 
In this interpretation, three
branes are points on the boundary of the disc and correspond to the
three fixed boundary conditions $A,B,C$. The other three describe
mixed boundary conditions $AB,BC,AC$ and are represented as lines (see
fig.\ \ref{pottsgeometry}).

The remaining two boundary conditions can be constructed as twisted
branes. 
They are labeled by pairs $[L,\pm]$ where $L=0,\dots ,3$
is an integer coming from the numerator part, and the sign $\pm $
comes from the twisted $U
(1)$. Selection and identification rules leave us with the two
boundary conditions $[0,+]\sim [3,-]$ and $[1,-]\sim [2,+]$.
In our geometric picture
the brane $[0,+]$ appears as point-like object in the center of the
disc, and the brane $[1,-]$ is a two-dimensional disc placed at the
origin (see fig.\ \ref{pottsgeometry}). 
They are the `free' and the `new' boundary condition, respectively.
Table~\ref{potts} gives an overview of boundary conditions in this
particular model.
\smallskip

Now, we want to apply our rule~\eqref{bcflow} to determine RG flows.
We first observe that the rule does not describe flows starting from a
single boundary condition. Instead, we shall analyze all possible
flows for superpositions of two boundary conditions. In all these
cases the boundary spin triggering the flow is $S=1$.

We start with untwisted branes. Applying the rule~\eqref{bcflow} for
$\alpha = [L=0,L'=1]$, we find the flow
\begin{eqnarray*}
\makebox[0pt][r]{$ A\oplus B \ =\ [0,0] \oplus  [0,2]$}&
\longrightarrow &\makebox[0pt][l]{$ [1,1] \ =\ AB \ \ .$}  \\
\parbox{1cm}{\makebox[1cm]{{\pottsAundB}}} \quad &\longrightarrow &\quad  \parbox{1cm}{\makebox[1cm]{{\pottsAB}}} 
\end{eqnarray*}
As one could already infer from symmetry arguments, there are also
the flows $B\oplus C\to BC$ and $A\oplus C\to
AC$. Starting instead with $\alpha = [1,2]$ we find
\begin{eqnarray*}
\makebox[0pt][r]{$ AB\oplus BC\ =\ [1,1]\oplus [1,3]$}&
\longrightarrow& \makebox[0pt][l]{$[0,2]\oplus [2,2]\
=\ B\oplus AC\ \ ,$} \\
\parbox{1cm}{\makebox[1cm]{{\pottsABundBC}}}\quad  &\longrightarrow &\quad  \parbox{1cm}{\makebox[1cm]{{\pottsBundAC}}} 
\end{eqnarray*}
analogous results can be obtained for permutations of the letters $A,B,C$.
 
Choosing $\alpha = [0,-]$ in \eqref{bcflow} yields the flow
\begin{eqnarray*}
\makebox[0pt][r]{$ 2\cdot F \ =\ 2\cdot [0,+]$}&
\longrightarrow &  \makebox[0pt][l]{$[1,-]\ =\ N \ \ .$}\\
\parbox{1cm}{\makebox[1cm]{{\pottsfreefree}}} \quad  & \longrightarrow & \quad  \parbox{1cm}{\makebox[1cm]{{\pottsnew}}} 
\end{eqnarray*}
If we set $\alpha = [1,+]$, the resulting flow is
\begin{eqnarray*}
\makebox[0pt][r]{$ 2\cdot N \ =\ 2\cdot [1,-] $}& \longrightarrow & \makebox[0pt][l]{$[0,+]\oplus
[2,+]\ =\ F \oplus N \ \ .$} \\
\parbox{1cm}{\makebox[1cm]{{\pottsnewnew}}} \quad  & \longrightarrow & \quad  \parbox{1cm}{\makebox[1cm]{{\pottsfreenew}}}
\end{eqnarray*}
These are all flows provided by the rule~\eqref{bcflow} for
superpositions of two boundary conditions. The field responsible for
the flows comes from the coset sectors $\cH ^{(0,\pm 2)}$ and has
conformal weight $h=2/3$. This can be concluded from
our general prescription in section~3 (see eq.\ \eqref{list}).
\medskip 

We now turn to the description of the Potts model as diagonal $su (2)$ coset,
\[
\frac{\widehat{su} (2)_{3}\oplus \widehat{su} (2)_{1}}{\widehat{su} (2)_{4}}
\]
where the modular invariant is obtained from charge-conjugated modular
invariants in the numerator, the denominator $su (2)_{4}$ contributes
a $D_{4} $ modular invariant. 
The perturbing field comes from the adjoint sector $[0,0,2]$ and has
conformal weight $h=2/3$.

We find four
boundary conditions $L_{1}=0,1,2,3$ in the $su (2)_{3}$ part and two boundary
conditions $L_{2}=0,1$ in the $su (2)_{1}$ part. The $su (2)_{4}$ part has
a $D_{4}$ modular invariant. There are
four boundary conditions which we label by $L'=0,1,2+,2-$. The
coefficients of the corresponding boundary states in terms of
Ishibashi states can be found e.g.\ in \cite{Behrend:1999bn}.

Identification and selection rules leave us with eight boundary
conditions for the 3-state Potts model. They are given in
table~\ref{potts}.
Applying our rule, we observe first that we find the same flows
involving superpositions of two boundary conditions that we discussed
in the parafermion construction. In addition we find flows relating
`free' and `new' boundary conditions with the others, namely (for
superpositions of maximally three boundary conditions):
\begin{equation*}
\renewcommand{\arraystretch}{2}
\begin{array}{ccccccl}
\parbox{1cm}{\makebox[1cm]{{\pottsfree}}} \ & = &\ F\ & \longrightarrow\ & A\ &=&\ \parbox{1cm}{\makebox[1cm]{{\pottsA}}} \\[1mm]
\parbox{1cm}{\makebox[1cm]{{\pottsfree}}} \ & = &\ F\ & \longrightarrow\ & AB\ &=&\ \parbox{1cm}{\makebox[1cm]{{\pottsAB}}}  \\[1mm]
\parbox{1cm}{\makebox[1cm]{{\pottsnew}}} \ & = &\ N \ & \longrightarrow\ & AB\ &=&\ \parbox{1cm}{\makebox[1cm]{{\pottsAB}}}  \\[1mm]
\parbox{1cm}{\makebox[1cm]{{\pottsnew}}} \ & = &\ N \ & \longrightarrow\ & AC\ \oplus\ B\
&=&\ \parbox{1cm}{\makebox[1cm]{{\pottsBundAC}}} \\[3mm]
\parbox{1cm}{\makebox[1cm]{{\pottsfreefree}}} \ &=&\ 2\cdot F\ & \longrightarrow\ & AB\ &=&\ \parbox{1cm}{\makebox[1cm]{{\pottsAB}}}  \\[1mm]
\parbox{1cm}{\makebox[1cm]{{\pottsnewnew}}} \ &=&\ 2\cdot N\ & \longrightarrow\ & AC\ \oplus\ B\
&=&\ \parbox{1cm}{\makebox[1cm]{{\pottsBundAC}}} \\[3mm]
\parbox{1cm}{\makebox[1cm]{{\pottsABC}}} \ & = &\ A\ \oplus\ B\ \oplus\ C\ & \longrightarrow\ & F\ &=&\ \parbox{1cm}{\makebox[1cm]{{\pottsfree}}} \\[1mm]
\parbox{1cm}{\makebox[1cm]{{\pottsABC}}} \ & = &\ A\ \oplus\ B\ \oplus\ C\ & \longrightarrow\ & N\ &=&\ \parbox{1cm}{\makebox[1cm]{{\pottsnew}}} \\[1mm]
\parbox{1cm}{\makebox[1cm]{{\pottsABundBCundCA}}} \ & = &\ AB\ \oplus\ BC\ \oplus\ AC\ &
\longrightarrow\ & N\ &=&\ \parbox{1cm}{\makebox[1cm]{{\pottsnew}}} \\[1mm]
\parbox{1cm}{\makebox[1cm]{{\pottsABundBCundCA}}} \ & = &\ AB\ \oplus\ BC\ \oplus\ AC\ &
\longrightarrow\ & F\ \oplus\ N\ &=&\
\parbox{1cm}{\makebox[1cm]{{\pottsfreenew}}} 
\end{array}
\end{equation*}
\medskip 

Let us finally discuss the construction of the Potts model as
\[
\frac{\widehat{su} (3)_{1}\oplus \widehat{su} (3)_{1}}{\widehat{su} (3)_{2}}
\]
coset. Its sectors are labeled by three $su (3)$ weights 
\[
[(l_{1},l_{2}),(m_{1},m_{2}),(l'_{1},l'_{2})]
\]
where $l_{i},m_{i},l'_{i}$ are non-negative integers (Dynkin labels)
obeying
\begin{gather*}
0\leq l_{1}+l_{2}\leq 1\quad ,\quad 0\leq m_{1}+m_{2}\leq 1\quad,
\quad 0\leq l'_{1}+l'_{2}\leq 2\\
2 (l_{1}+m_{1}-l'_{1})+l_{2}+m_{2}-l'_{2}\ =\ 0\ \mod 3 \ \ .
\end{gather*}
The sectors are identified according to the field identification
\begin{multline*}
[(l_{1},l_{2}),(m_{1},m_{2}),(l'_{1},l'_{2})]\ \sim \\
\sim \ [(1-l_{1}-l_{2},l_{1}),(1-m_{1}-m_{2},m_{1}),(2-l'_{1}-l'_{2},l'_{1})]\
\ .
\end{multline*}
What remains are 6 sectors. According to the standard Cardy
construction, these give rise to 6 boundary conditions which are
listed in table~\ref{potts} along with their g-factors. Before we go to
construct the remaining two boundary conditions, we want to look for
RG flows.

Let us start with the boundary condition $AB$ and exhibit what flows
are `predicted' by \eqref{bcflow}. We choose the perturbation $\sspin =
((0,0),(0,1))$ and find the flow
\begin{eqnarray*}
\makebox[0pt][r]{$ AB\ =\ [(0,0),(1,0),(1,0)]$} &  \longrightarrow & \makebox[0pt][l]{$[(0,0),(0,0),(0,0)]\
=\ A\ \ .$}\\
\parbox{1cm}{\makebox[1cm]{{\pottsAB}}} \quad & \longrightarrow & \quad \parbox{1cm}{\makebox[1cm]{{\pottsA}}} 
\end{eqnarray*}
The spin $\sspin = ((0,1),(0,0))$ leads to
\begin{eqnarray*}
\makebox[0pt][r]{$ AB\ =\ [(0,0),(1,0),(1,0)]$} & \longrightarrow & \makebox[0pt][l]{$[(0,0),(0,1),(2,0)]\ =\
B\ \ .$} \\
\parbox{1cm}{\makebox[1cm]{{\pottsAB}}} \quad  & \longrightarrow & \quad \parbox{1cm}{\makebox[1cm]{{\pottsB}}} 
\end{eqnarray*}
Analogously, we find $BC\to B$, $BC\to C$ and $AC\to A$, $AC\to
C$. These constitute all flows from single boundary conditions described by
the rule. 
For a superposition of two boundary conditions we find flows
of the form 
\begin{eqnarray*}
AC\oplus B \quad  &\longrightarrow & \quad \ \ A  
\makebox[0cm]{\hspace*{0.7cm} .} \\
\parbox{1cm}{\makebox[1cm]{{\pottsBundAC}}} \quad & \longrightarrow & \quad \parbox{1cm}{\makebox[1cm]{{\pottsA}}} 
\end{eqnarray*}
\smallskip

The two remaining boundary conditions can be obtained from twisted
gluing conditions using an automorphism which interchanges the two
Dynkin labels of the $su (3)$ theories. In the $su (3)_{1}$ there is
only one sector left invariant under this automorphism, in the $su
(3)_{2}$ theory there are two. In total we find two twisted boundary
conditions
\[
[0,0,0;\omega ]\quad \text{and}\quad [0,0,1;\omega ]\ \ ,
\]
there are no selection or identification rules in this example.
We can calculate their g-factors (see table~\ref{potts}) and
identify the two boundary conditions as the `new' and the `free'
boundary condition, respectively.
\begin{figure}[!t]
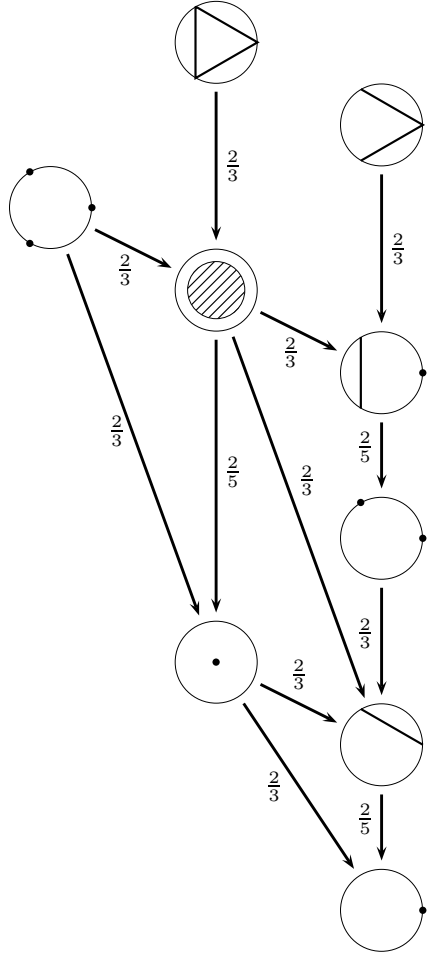

\begin{center}
{\pottsprocesses}
\end{center}
\caption{\label{pottsprocesses} 
Some of the boundary RG flows found in the
3-states Potts model. The vertical ordering of the configurations is done
according to the g-factors. The conformal weight of the field responsible
for a flow is quoted. 
}
\end{figure}

Again, we want to investigate what flows are described by the
rule~\eqref{bcflow}. Let us start with the `new' boundary condition and
try the perturbation $\sspin = ((1,0),(0,0))$. This leads to
\begin{eqnarray*}
\makebox[0pt][r]{$N\ =\   [0,0,0;\omega ] $}& \longrightarrow &
\makebox[0pt][l]{$[0,0,1;\omega ]\ =\ 
F\ \ .$}\\
\parbox{1cm}{\makebox[1cm]{{\pottsnew}}} \quad & \longrightarrow & \quad \parbox{1cm}{\makebox[1cm]{{\pottsfree}}} 
\end{eqnarray*}

We can identify the field that drives the described flows. From our
general prescription~\eqref{list}, we conclude that the perturbing
field is $[(0,0),(0,0),(1,1) ]$
and has conformal weight $h=2/5$.
\medskip 

Let us compare our results with the work of Affleck et
al.\cite{Affleck:1998nq} (see also \cite{Affleck:2000ws}). 
They find several flows driven by fields of
conformal weight $h=2/3$ and $h=2/5$. The flows they find are \textit{all}
reproduced by our rule. For single boundary conditions we find exact
coincidence, for superpositions our rule suggests further flows that
have not been analyzed in \cite{Affleck:1998nq}.
\smallskip

Figure~\ref{pottsprocesses} summarizes part of the results for
boundary RG flows in the 3-states Potts model obtained by the
rule~\eqref{bcflow}. The complete results can be found in appendix~B.

\subsection[$N=2$ Minimal Models]{$\boldsymbol{N=2}$ Minimal Models}
As last example we choose the supersymmetric parafermion theories, the
$N=2$ minimal series. They can be constructed as cosets $\widehat{su}
(2)_{k}\oplus \widehat{u}_{4}/\widehat{u}_{2k+4}$. The sectors of the
theory are labeled by triples
$[l_{1},l_{2},l']$ where\footnote{Usually the sectors are labeled
by the triples $(l,m,s)$ where $l$ corresponds to $l_{1}$, $m$ to $l'$ and $s$
corresponds to $l_{2}$. We choose a different order here and
put all labels that belong to the numerator theory to the front.}  
$l_{1}=0,\dots ,k$ and $l'$
is a $(2k+4)$-periodic integer with standard range $l'=-k-1,\dots
,k+2$. The third label $l_{2}$ can take the values $l_{2}=-1,0,1,2$. Selection
rules force $l_{1}+l_{2}+l'$ to be even and the triples $[l_{1},l_{2},l']$ and
$[k-l_{1},l_{2}\pm 2,l'\pm (k+2)]$ label the same sectors.

The discussion of boundary conditions is analogous to the parafermion case
(section~4.5). There are untwisted boundary conditions (A-branes), labeled by
triples $[L_{1},L_{2},L']$, and twisted ones
(B-branes)~\cite{Maldacena:2001ky}. As many results can be directly
translated from the parafermion case, we are not going to repeat the
whole picture here, but restrict our discussion now
to untwisted, `even' ($L_{2}=0,2$) boundary conditions. 
\begin{figure}[!t]
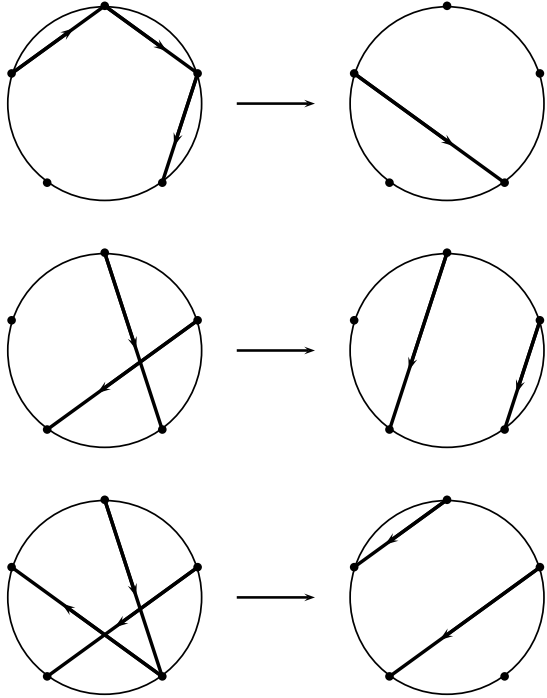

\begin{center}
\scalebox{.65}{\supminmodprocI}
\scalebox{.65}{\supminmodprocII}
\scalebox{.65}{\supminmodprocIII}
\end{center}
\caption{\label{supminmodprocI}Some flows in the $N=2$ minimal model
for $k=3$.}
\end{figure}

Geometrically they can be displayed by straight, oriented lines
(orientation depends on the label $L_{2}$)
stretched between $k+2$ special punctures on a
disc~\cite{Maldacena:2001ky}. The smallest
lines connecting two neighboring points have a label $L_{1}=0,k$.

\begin{figure}
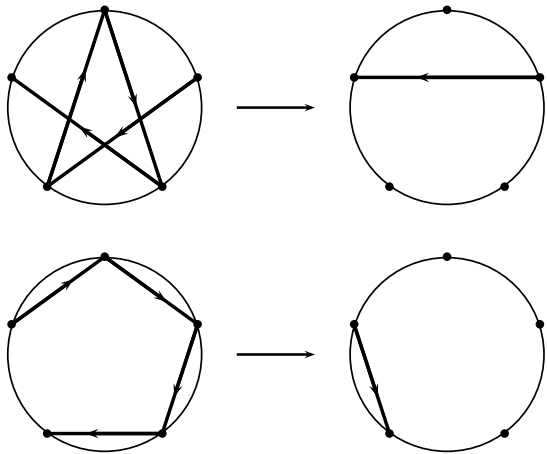

\begin{center}
\scalebox{.65}{\supminmodprocIV}
\scalebox{.65}{\supminmodprocV}
\end{center}
\caption{\label{supminmodprocII}Flows~\eqref{supminmodflow} in the
$N=2$ minimal model for $k=3$: the first one is obtained by setting
$S=k,L_{1}=1$, the second one by setting $S=k,L_{1}=0$.}
\end{figure}

Let us see what flows are described by the rule~\eqref{bcflow}. Let us
choose $L_{1},L_{2},L',S$ s.t.\ $L_{1}+L_{2}+L'+S$ is even. Then we
find the flow 
\begin{multline}\label{supminmodflow}
[L_{1},L_{2},L'-S]\oplus [L_{1},L_{2},L'-S+2]\oplus \dots \oplus
[L_{1},L_{2},L'+S]\\  \longrightarrow\
\bigoplus_{J}N^{(k)}_{SL_{1}}{}^{J}[J,L_{2},L']\ \ .
\end{multline}
We show some examples for $k=3$ in
figures~\ref{supminmodprocI} and~\ref{supminmodprocII}.
\smallskip

At the end of this section, 
we want to show how these results can be used to determine
the group of brane charges in the $N=2$ minimal models. We assign
charges $Q_{[L_{1},L_{2},L']}$ to the boundary conditions 
s.t.\ they are conserved during
RG flows. From the rule we see immediately that all charges can be
expressed by the charges $Q_{[0,L_{2},L']}$ 
of boundary conditions with $L_{1}=0$ (just
set $L_{1}$ to zero in the flow~\eqref{supminmodflow}). 
This means that for the even
untwisted boundary conditions we have at most the charge group $\ZZ
^{2k+4}$, one copy of $\ZZ $ for every boundary condition with
$L_{1}=0$. Now, the rule implies more constraints on the charges.
It turns out that in the end we remain with the relations
\begin{subequations}
\begin{align}\label{firstrelation}
Q_{[0,0,L'-k-1]}+Q_{[0,0,L'-k+1]}+\dots +Q_{[0,0,L'+k+1]}\ & =\ 0\\ 
\label{secondrelation}
Q_{[0,0,L']}\ & =\ - Q_{[0,2,L']}\ \ .
\end{align}
\end{subequations}
To find~\eqref{firstrelation}, it is sufficient to consider the flows with
$S=k$ and $L_{1}=1$ in~\eqref{supminmodflow} (see e.g.\ the first flow
of figure~\ref{supminmodprocII}).
The second relation~\eqref{secondrelation} results from flows 
with $S=k$ and $L_{1}=0$ (see the second flow of
figure~\ref{supminmodprocII}) combined with~\eqref{firstrelation}. 
One can then show that other flows do not give any further constraints.

To summarize we find that the charge group of even untwisted branes is
$\ZZ ^{k+1}$. This result has already been obtained in~\cite{Maldacena:2001ky}.
It coincides with the computation of RR-charges in~\cite{Lerche:2000iv}.

\section{Conclusions}
In this work we presented a proposal for a simple rule for boundary
RG flows in coset models. A specialized version of the rule for
untwisted Cardy boundary conditions had been announced
earlier~\cite{Fredenhagen:2002qn}. Evidence for the rule comes from
three directions. First, the rule is in concordance with perturbative
results. Second, there are strong arguments that 
all flows described by the rule satisfy the general `g-theorem' of
Affleck and Ludwig. Third, in a number of examples, including
non-trivial exceptional coset constructions, the rule is able to
reproduce flows that have been obtained by different means.

In the last section of the paper, we presented the broad range of
application of the rule in different coset constructions. Having
a rich knowledge of possible boundary RG flows in particular models,
we can start to determine quantities that are invariant under 
RG flows. Interpreting the RG flows as dynamical processes between
brane configurations in string theory, the determination of invariants 
leads to the computation of D-brane charge groups. For WZNW models such
a computation has been performed 
in~\cite{Fredenhagen:2000ei, Bouwknegt:2002bq} using the
rule of Affleck and Ludwig. We showed how the generalized rule can
be used to determine brane charges in $N=2$ minimal models. The
same method can be applied to other coset models, in particular to
other Kazama-Suzuki models and compared to the RR-charges that have
been calculated in~\cite{Lerche:2000iv}.

Let us mention two open issues that one could consider in the
future. In the perturbative regime we see a lot more flows than are
described by our rule. What happens to them when we take the level to
finite values? Another problem which is almost unexplored concerns
boundary RG flows for non-maximally symmetric boundary
conditions. In the perturbative regime, we have some (though very
limited) informations from non-symmetric solutions of the
effective action found in~\cite{Alekseev:2002rj}. It would be
interesting to study these boundary conditions and their relevance for
brane charges further.

\enlargethispage{.5cm}
\subsection*{Acknowledgements}
I would like to thank
A.\  Cappelli, G.\ D'Appollonio, K.\ Graham, A.\ Recknagel, T.\ Quella,
C.\ Schweigert and especially V.\ Schomerus for their comments 
and useful and stimulating discussions.

\begin{appendix}
\section{Compatibility with g-conjecture}
In this appendix we want to show that the
inequality~\eqref{inequality} which describes the compatibility of our
proposed rule with the g-conjecture is satisfied when we take 
the levels to be large. For convenience, let us rewrite the inequality
here and introduce the abbreviations $g_{L}$ and $g_{R}$ for left and
right hand side of the relation,
\begin{equation}\label{app:inequality}
g_{L}\ :=\ \sum_{S'}b_{SS'}\frac{S^{\Lie{h}}_{S'0}}{S^{\Lie{h}}_{00}}\
> \ \frac{S^{\Lie{g}}_{S0}}{S^{\Lie{g}}_{00}}\ =:\ g_{R}\ \ .
\end{equation}
We shall expand both sides of the relation to the order $1/k^{2}$ where
$k$ is the level which is sent to large values. Note that we not
necessarily have to take all levels to be large as long as the
representation $S$ is trivial w.r.t.\ the algebras that stay at small
levels.
\smallskip

Let us consider the simplest case when $\Lie{g}$ and $\Lie{h}$ are
simple Lie algebras with level $k$ and $k'=kx_{e}$, respectively. Here,
the integer $x_{e}$ denotes the embedding index of the embedding
$\Lie{h}\hookrightarrow\Lie{g}$. When we want to expand the
expressions in~\eqref{app:inequality}, we can make use of the
formula 
\begin{equation}\label{expansionS}
\frac{S^{\Lie{g}}_{S0}}{S^{\Lie{g}}_{00}}\ =\ \dim
(V^{S})\Bigl(1-\frac{\pi^{2}}{6 k^{2}}g^{\vee }C_{S}
\Bigr) + \mc{O}( 1/k^{3})
\end{equation}
which can be found e.g.\ in~\cite[eq.\
(13.175)]{FrancescoCFT}. Here, $g^{\vee }$ is the dual Coxeter number
of $\Lie{g}$, $V^{S}$ is the representation space of
the representation labeled by $S$, and $C_{S}$ is the quadratic Casimir,
\[
C_{S}\ =\ \frac{1}{\dim V^{S}} \tr _{V^{S}} \bigl(T_{\mu}T^{\mu }\bigr)\ \ .
\]
$T_{\mu }$ are the generators of the Lie algebra $\Lie{g}$. To relate
the Casimir of a $\Lie{g}$-representation with that of a
representation of $\Lie{h}$ we use the formula
\begin{equation}\label{Casimirrelation}
\tr_{V^{S}} \bigl(T_{\mu }T^{\mu }\bigr)\ =\ \frac{1}{x_{e}}\frac{\dim
\Lie{g}}{\dim \Lie{h}}\, \tr _{V^{S}} \bigl(T'_{m}T'^{m}\bigr)
\end{equation}
where the $T'_{m}$ are 
the generators of $\Lie{h}$ embedded in $\Lie{g}$. 

Now we are prepared to check the inequality~\eqref{app:inequality}. 
We expand the l.h.s.\ according to the formula~\eqref{expansionS} and
obtain
\[
g_{L} \ =\ \sum_{S'}b_{SS'} \dim (V^{S'})\Bigl(1-\frac{\pi
^{2}}{6k'^{2}}g_{\Lie{h}}^{\vee }C_{S'} \Bigr) + \mc{O} (1/k'^{3})\ \ .
\]
After some manipulations and applying the
relation~\eqref{Casimirrelation} we arrive at
\[
g_{L}\ =\ \dim (V^{S})\Bigl(1-\frac{\pi ^{2}}{6
k'^{2}}g_{\Lie{h}}^{\vee } \frac{x_{e}\dim \Lie{h}}{\dim \Lie{g}}C_{S}
\Bigr) + \mc{O} (1/k'^{3})\ \ .
\]
Inserting $k'=kx_{e}$ and using the fact that $\dim \Lie{g}>\dim
\Lie{h}$ as well as $g^{\vee }>g_{\Lie{h}}^{\vee }/x_{e}$ we can
finally conclude that $g_{L}>g_{R}$ in the order $1/k^{2}$.
\smallskip

The proof for semi-simple Lie algebras can be done essentially 
along the same lines. Let us sketch the procedure in the example of
a coset model of the form $\Lie{g}^{(1)}_{k_{1}}\oplus \cdots \oplus
\Lie{g}^{(n)}_{k_{n}} /\Lie{h}_{k'}$ where the level of the
denominator is determined by $k'=k_{1}x_{1}+\cdots+k_{n}x_{n}$. Using
the expansion formula~\eqref{expansionS}, we reduce our problem to
proving the inequality 
\begin{equation}\label{reducedproblem}
\sum_{i=1}^{n} \frac{g^{\vee}_{i}}{k_{i}^{2}}\, \tr  \bigl(T_{i}^{2}\bigr)\ >\
\frac{g_{h}^{\vee }}{k'^{2}}\, \tr \bigl({\textstyle \sum}\,  T_{i}'\bigr)^{2}\ \ .
\end{equation}
The left hand side can be estimated from below as before,
\begin{equation}\label{estimate}
\sum_{i=1}^{n} \frac{g^{\vee}_{i}}{k_{i}^{2}}\, \tr \bigl(T_{i}^{2}\bigr)\ >\
\frac{1}{\sum k_{j }x_{j}}\sum_{i}\frac{g_{h}^{\vee }}{k_{i}x_{i}}\, \tr
\bigl( T_{i}'^{2}\bigr) \ \ . 
\end{equation}
We show now that this estimate is good enough to
prove~\eqref{reducedproblem}, i.e.\ that the difference 
\[
\eta \ :=\ \sum_{i}\frac{\sum_{j}k_{j}x_{j}}{k_{i}x_{i}}\,\tr \bigl(
T_{i}'^{2}\bigr)
- \tr \bigl( {\textstyle \sum}\, T'_{i} \bigr) ^{2}
\]
is positive. By introducing
$a_{i}:=\sqrt{k_{i}x_{i}/\sum_{j}k_{j}x_{j}}$ we can rewrite $\eta $
in a manifestly positive form,
\[
\eta\ =\ \sum_{i=1}^{n-1}\tr \Bigl[\frac{a_{i}}{1+a_{n}}
(T'_{1}+\cdots + T'_{n-1})+\frac{a_{i}}{a_{n}}T'_{n}-\frac{1}{a_{i}}T'_{i}
 \Bigr]^{2} \ \ ,
\]
which completes the proof.

\section{Tables for flows in specific models}
\subsection{Critical Ising model}
Boundary conditions: $0$ (free), $+$ (spin up), $-$ (spin down).
\begin{itemize}
\item {\bf Coset realization} $\displaystyle \frac{su (2)_{1}\oplus su
(2)_{2}}{su (2)_{3}}$: Perturbing field has $h=1/2$.\\ 
Flows resulting from~\eqref{bcflow} starting from
\begin{itemize}
\item a {\bf single} boundary condition
\begin{align*}
0\ &\longrightarrow \ +\\
&\longrightarrow \ -
\end{align*}
\item a superposition of {\bf two} boundary conditions
\newlength{\tmpA}
\settowidth{\tmpA}{$0$}
\begin{align*}
\makebox[\tmpA][r]{$+\oplus -$}\ &\longrightarrow\ 0\\
&\longrightarrow\ +\\
&\longrightarrow\ -
\end{align*} 
\end{itemize}
\end{itemize}

\subsection{Tricritical Ising model}
Boundary conditions:
\begin{center}
\begin{tabular}{ccrcl}
Symbol & \parbox{2.72cm}{Conf.\ weight $h$ of corresp.\ field} &
\multicolumn{3}{c}{g-factor}\\ \hline 
$+$     & $3/2$  & $a$            & $\approx $ & $.5127$ \\
$-$     & $0$    & $a$            & $\approx $ & $.5127$ \\
$0$     & $7/16$ & $a\sqrt{2}$    & $\approx $ & $.7251$ \\
$0\! +$ & $3/5$  & $b/a\sqrt{2}$  & $\approx $ & $.8296$ \\ 
$-\! 0$ & $1/10$ & $b/a\sqrt{2}$  & $\approx $ & $.8296$ \\
$d$     & $3/80$ & $b/a$          & $\approx $ & $1.173$ 
\end{tabular}
\end{center}
Here, $\displaystyle a^{4}=\frac{5-\sqrt{5}}{40}\ ,\
b^{2}=\frac{5+\sqrt{5}}{2}$.

\begin{itemize}
\item {\bf Coset realization} $\displaystyle \frac{su (2)_{2}\oplus su
(2)_{1}}{su (2)_{3}}$: Perturbing field has $h=3/5$.\\
Flows resulting from~\eqref{bcflow}  starting from
\settowidth{\tmpA}{$2\cdot d \oplus 0$}
\newlength{\tmpB}\settowidth{\tmpB}{${0\! +}\oplus {-\! 0}$}
\newlength{\tmpC}\settowidth{\tmpC}{$2\cdot {-\! 0}\oplus +$}
\newlength{\tmpD}\settowidth{\tmpD}{${-\! 0}$}
\newlength{\tmpE}\settowidth{\tmpE}{$2\cdot {0\! +}\oplus -$}
\newlength{\tmpF}\settowidth{\tmpF}{${0\! +}$}
\begin{itemize}
\item a {\bf single} boundary condition
\begin{align*}
\makebox[\tmpA][r]{$d$}\ &\lra \ \makebox[\tmpB][l]{$0$}         &
\makebox[\tmpC][r]{$0\! +$}\ &\lra \ \makebox[\tmpD][l]{$0$} &
\makebox[\tmpE][r]{$-\! 0$}\ &\lra \ \makebox[\tmpF][l]{$0$} \\
   &\lra \ +\oplus - &     &\lra \ + &     &\lra \ -
\end{align*}
\item a superposition of {\bf two} boundary conditions
\begin{align*}
\makebox[\tmpA][r]{$0\oplus d$}\ &\lra \ d           &
\makebox[\tmpC][r]{$-\oplus 0\! +$}\ &\lra \ -  &
\makebox[\tmpE][r]{$+\oplus -$}\ &
\lra \ +  \\
           &\lra \ 0           &             &\lra \ -\! 0 &            &
\lra \ 0\! + \\
           &\lra \ +\oplus -   &             &\lra \ d  &            &
\lra \ d  \\
           &\lra \ {0\! +}\oplus -\! 0 &             &\lra \ 0  &            &
\lra \ 0
\end{align*}
\item a superposition of {\bf three} boundary conditions
\begin{align*}
2\cdot d \oplus 0\ &\lra \ d           & 2\cdot -\! 0\oplus +\ &\lra \ d &
2\cdot {0\! +}\oplus -\ &\lra \ d \\
                   &\lra \ -\! 0\oplus 0\! + &
&\lra \ -\! 0&
                   &\lra \ 0\! +
\end{align*}
\end{itemize}
\item {\bf Coset realization} $\displaystyle \frac{(E_{7})_{1}\oplus
(E_{7})_{1}}{(E_{7})_{2}}$: Perturbing field has $h=1/10$.\\ 
Flows resulting from~\eqref{bcflow} starting from
\settowidth{\tmpA}{$+\oplus -\oplus 2\cdot {0\! +}\oplus 2\cdot {-\! 0}$}
\settowidth{\tmpB}{${0\! +}$}
\settowidth{\tmpC}{$2\cdot 0\oplus 2\cdot d$}
\settowidth{\tmpD}{${0\! +}$}
\begin{itemize}
\item a {\bf single} boundary condition
\begin{align*}
d\ &\lra \ +\\
   &\lra \ -
\end{align*}
\item a superposition of {\bf two} boundary conditions
\begin{align*}
\makebox[\tmpA][r]{${{0\! +}\oplus {-\! 0}}$}\ &\lra \
\makebox[\tmpB][l]{$0$} & \makebox[\tmpC][r]{$0\oplus d$}\ &\lra \ 0\! +  \\
             &         &            &\lra \ -\! 0  
\end{align*}
\item a superposition of {\bf four} boundary conditions
\begin{align*}
\makebox[\tmpA][r]{$+\oplus -\oplus {0\! +}\oplus {-\! 0}$}\ &\lra \
\makebox[\tmpB][l]{$d$} & 2\cdot 0\oplus 2\cdot d \
&\lra \ \makebox[\tmpD][l]{$0$} \\
                             &\lra \ + & & \\
                             &\lra \ - & &
\end{align*} 
\item a superposition of {\bf six} boundary conditions
\begin{align*}
+\oplus -\oplus 2\cdot {0\! +}\oplus 2\cdot {-\! 0}\ &\lra \ 0\! + & 2\cdot 0\oplus
4\cdot d\ &\lra \ \makebox[\tmpD][l]{$d$} \\
                                           &\lra \ -\! 0 & &
\end{align*}
\end{itemize}  
\end{itemize}
\subsection{Three-state Potts model}
Boundary conditions: $A,B,C,AB,BC,AC,F,N$ (see table~\ref{potts} on
page~\pageref{potts}). Note that the model has a $\ZZ _{3}$-symmetry which
acts on the boundary conditions as $A\to B\to C\to A,AB\to BC\to AC\to
AB$, the boundary conditions $F$ and $N$ are fixed. We shall only write
out flows modulo this symmetry, e.g.\ the flow $A\oplus
B\longrightarrow AB$ stands also for $B\oplus C\longrightarrow BC$ and
$C\oplus A\to AC$.

\begin{itemize}
\item {\bf Coset realization} $\displaystyle \frac{su (2)_{1}\oplus su
(2)_{3}}{su(2)_{4}}$: Perturbing field has $h=2/3$.\\
Flows resulting from~\eqref{bcflow} starting from
\begin{itemize}
\item a {\bf single} boundary condition
\begin{align*}
F\ &\lra \ BC & N\ &\lra \ BC        \\
   &\lra \ A  &    &\lra \ A\oplus BC 
\end{align*}
\item a superposition of {\bf two} boundary conditions
\begin{align*}
2\cdot F\ &\lra \ N  & 2\cdot N\ &\lra \ F\oplus N  & A\oplus B\ &\lra
\ AB \\
          &\lra \ BC &           &\lra \ A\oplus BC & AB\oplus AC&\lra
\ A\oplus BC
\end{align*}
\item a superposition of {\bf three} boundary conditions
\begin{align*}
A\oplus B\oplus C\ &\lra \ N  & AB\oplus BC\oplus AC\ &\lra \ F\oplus
N  \\
                   &\lra \ F  &                       &\lra \ N
   \\
                   &\lra \ AB &                       &\lra \ A\oplus
BC \\
                   &\lra \ A  &                       &\lra \ AB
   \\[2mm]
3\cdot F\ &\lra \ N & 3\cdot N\ &\lra \ F\oplus N \\
          &\lra \ F &           &\lra \ N
\end{align*}
\item a superposition of {\bf six} boundary conditions
\begin{align*}
2\cdot A\oplus 2\cdot B\oplus 2\cdot C \ &\lra \ N\\
2\cdot AB\oplus 2\cdot BC\oplus 2\cdot AC \ &\lra \ F\oplus N\\
\end{align*}
\end{itemize}
\item {\bf Coset realization} $\displaystyle \frac{su (2)_{3}}{u
(1)}$: Perturbing field has $h=2/3$.\\ 
Flows resulting from~\eqref{bcflow} starting from
\settowidth{\tmpA}{$2\cdot AB \oplus BC \oplus AC$}
\settowidth{\tmpB}{$A\oplus BC$}
\settowidth{\tmpD}{$F\oplus N$}
\begin{itemize}
\item a superposition of {\bf two} boundary conditions
\begin{align*}
\makebox[\tmpA][r]{$A\oplus B$}\   &\lra \ AB         & 2\cdot F\ &\lra \ N \\
AB\oplus AC\ &\lra \ A\oplus BC & 2\cdot N\ &\lra \ F\oplus N
\end{align*} 
\item a superposition of {\bf three} boundary conditions
\begin{align*}
\makebox[\tmpA][r]{$A\oplus B\oplus C$}\    &\lra \ AB         & 3\cdot F\ &\lra \ N \\
AB\oplus BC\oplus AC\ &\lra \ A\oplus BC & 3\cdot N\ &\lra \ F\oplus N
\end{align*}
\item a superposition of {\bf four} boundary conditions
\begin{align*}
2\cdot A\oplus B\oplus C\       &\lra \ \makebox[\tmpB][l]{$A$}  &
4\cdot F\ &\lra \ \makebox[\tmpD][l]{$F$}\\
2\cdot AB \oplus BC \oplus AC \ &\lra \ AB & 4\cdot N\ &\lra \ N
\end{align*}
\end{itemize}
\item {\bf Coset realization} $\displaystyle \frac{su (3)_{1}\oplus su
(3)_{1}}{su (3)_{2}}$: Perturbing field has $h=2/5$.\\ 
Flows resulting from~\eqref{bcflow} staring from
\settowidth{\tmpA}{$A\oplus 2\cdot BC$}
\settowidth{\tmpB}{$AC$}
\settowidth{\tmpC}{$F\oplus 2\cdot N$}
\settowidth{\tmpD}{$N$}
\begin{itemize}
\item a {\bf single} boundary condition
\begin{align*}
\makebox[\tmpA][r]{$AB$}\ &\lra \ \makebox[\tmpB][l]{$A$} &
\makebox[\tmpC][r]{$N$}\ &\lra \ \makebox[\tmpD][l]{$F$} \\
    &\lra \ B &    &
\end{align*}
\item a superposition of {\bf two} boundary conditions
\begin{align*}
\makebox[\tmpA][r]{$A\oplus BC$}\ &\lra \ AB &
\makebox[\tmpC][r]{$F\oplus N$}\ &\lra \ N \\
            &\lra \ AC &            &\lra \ F \\
            &\lra \ A  &            &         \\
            &\lra \ B  &            &         \\
            &\lra \ C  &            &
\end{align*}
\item a superposition of {\bf three} boundary conditions
\begin{align*}
A\oplus 2\cdot BC\ &\lra \ BC & F\oplus 2\cdot N\ &\lra \ N \\
                   &\lra \ AB & & \\
                   &\lra \ AC & &
\end{align*}
\end{itemize}
\end{itemize}
\end{appendix}

\end{document}